\documentclass[aip,graphicx ,
twocolumn,
reprint
]{revtex4-1}

\usepackage{graphicx,dcolumn,bm}
\usepackage{amsfonts,amsmath,amssymb,amsthm,amscd}
\usepackage[english]{babel}
\usepackage{physics}
\usepackage{needspace}
\usepackage{xcolor}
\usepackage{bm}
\usepackage{slashed}
\usepackage[export]{adjustbox}
\usepackage{tabularx}

\definecolor{linkblue}{HTML}{2e3092}
\definecolor{textblue}{HTML}{0208D8}
\usepackage[colorlinks=true, allcolors = linkblue]{hyperref}
\renewcommand{\vec}[1]{\mathbf{#1}}
\newcommand{\tem}{t_{\mathrm{em}}}
\newcommand{\gem}{\gamma_{\mathrm{em}}}
\newcommand{\cem}{\chi_{\mathrm{em}}}
\AtBeginDocument{\renewcommand{\natexlab}[1]{#1}}

\newcommand{\WCS}{W_{\mathrm{nCE}}}

\newcommand{\WBW}{W_{\mathrm{nBW}}}

\newcommand{\WCSBW}{W_{\mathrm{nCE,nBW}}}

\newcommand{\Ns}{N_{\mathrm{seed}}}
\newcommand{\Nsf}{N_{\mathrm{seed}}^{\mathrm{foc}}}
\newcommand{\atot}{a_*}
\newcommand{\Ptot}{P_*}
\newcommand{\aS}{a_{S}}
\newcommand{\ath}{a_{\mathrm{th}}}
\newcommand{\erad}{\epsilon_{\mathrm{rad}}}

\newcommand{\CORR}[1]{\textcolor{black}{#1}}

\begin{document}

\title{Testing strong-field QED with the avalanche precursor}

\author{\small A. A. Mironov}
\affiliation{Center for Theoretical Physics (CPHT), CNRS, \'{E}cole Polytechnique, Institut Polytechnique de Paris, 91128 Palaiseau, France}
\email{mironov.hep@gmail.com}

\author{\small S. S. Bulanov}
\affiliation{ Lawrence Berkeley National Laboratory, Berkeley, CA, USA}

\author{\small A. Di Piazza}
\affiliation{Department of Physics and Astronomy and Laboratory for Laser Energetics, University of Rochester, Rochester, NY, USA}

\author{\small M.  Grech}
\affiliation{ Laboratoire pour l'Utilisation des Lasers Intenses, CNRS, Sorbonne Université, CEA, Ecole Polytechnique, Institut Polytechnique de Paris, Palaiseau, France}

\author{\small L. Lancia}
\affiliation{ Laboratoire pour l'Utilisation des Lasers Intenses, CNRS, Sorbonne Université, CEA, Ecole Polytechnique, Institut Polytechnique de Paris, Palaiseau, France}

\author{\small S. Meuren}
\affiliation{ Laboratoire pour l'Utilisation des Lasers Intenses, CNRS, Sorbonne Université, CEA, Ecole Polytechnique, Institut Polytechnique de Paris, Palaiseau, France}

\author{\small J.  Palastro}
\affiliation{ University of Rochester, Laboratory for Laser Energetics, Rochester, NY, USA}

\author{\small C. Riconda}
\affiliation{ Laboratoire pour l'Utilisation des Lasers Intenses, Sorbonne Université, CNRS, CEA, Ecole Polytechnique, Institut Polytechnique de Paris, Paris, France}

\author{\small H. G. Rinderknecht}
\affiliation{ University of Rochester, Laboratory for Laser Energetics, Rochester, NY, USA}

\author{\small P. Tzeferacos}
\affiliation{ University of Rochester, Laboratory for Laser Energetics, Rochester, NY, USA}

\author{\small G. Gregori}
\affiliation{ Department of Physics,  University of Oxford, Parks Road, Oxford OX1 3PU, Oxford, UK}

\date{\today}

\begin{abstract}
A two-beam high-power laser facility is essential for the study of one of the most captivating phenomena predicted by strong-field quantum electrodynamics (QED) and yet unobserved experimentally: the avalanche-type cascade. 
In such a cascade, the energy of intense laser light can be efficiently transformed into high-energy radiation and electron-positron pairs. The future 50-petawatt-scale laser facility NSF OPAL will provide unique opportunities for studying such strong-field QED effects, as it is designed to deliver two ultra-intense, tightly focused laser pulses onto the interaction point.
In this work, we investigate the potential of such a facility for studying elementary particle and plasma dynamics deeply in the quantum radiation-dominated regime, and the generation of QED avalanches.
With 3D particle-in-cell simulations, we demonstrate that QED avalanche precursors can be reliably triggered under realistic laser parameters and layout (namely, focusing $f/2$, tilted optical axes, and non-ideal co-pointing) with the anticipated capabilities of NSF OPAL. 
We demonstrate that seed electrons can be efficiently injected into the laser focus by using targets of three types: a gas of heavy atoms, an overcritical plasma, and a thin foil.
A strong positron and high-energy photon signal is generated in all cases.
The cascade properties can be identified from the final particle distributions, which have a clear directional pattern.
At increasing laser field intensity, such distributions provide signatures of the transition, first, to the radiation-dominated interaction regime, and then to a QED avalanche. Our findings can also be used for designing related future experiments.
\end{abstract}


\maketitle 

\section{Introduction}
\label{sec:i}

The interaction of charged particles and photons with strong electromagnetic (EM) fields has attracted considerable interest for almost a century.\cite{Di_Piazza_2012,Gonoskov_2022,fedotov2023high,popruzhenko2023dynamics}  Even before the modern formulation of quantum electrodynamics (QED), the process of pair production under the action of a strong EM field, vacuum nonlinearities, and the attainability of the maximum possible EM field strength were studied and discussed.\cite{sauter.zp.1931,heisenberg.zp.1936} At some point, it was understood that the processes of photon emission by charged particles and electron-positron pair production by photons in strong EM fields exhibit properties that were not encountered in perturbative QED, moreover, the interplay between these processes and particle dynamics in EM fields gives rise to new phenomena.\cite{schwinger1951,reiss1962absorption,nikishov_jetp1964,brown1964interaction} \CORR{The initial studies of strong-field effects were mainly of theoretical interest, motivated by astrophysical applications.\cite{erber1966high,baier1968,ritus1985}}
However, high-intensity lasers, which became available after the introduction of the chirped-pulse-amplification technique,\cite{strickland1985compression} decisively changed the landscape of QED studies in the presence of strong EM fields, which are often referred to as Strong Field Quantum Electrodynamcis (SFQED). The flexibility of the laser EM field configuration and the ability of lasers to accelerate particles from underdense \cite{esarey.rmp.2009} and solid density \cite{daido.rpp.2012} plasmas and then make them interact with strong EM fields of that very laser or a second one has paved the way for a plethora of experimental proposals to study QED in strong EM fields.  

Though the interaction of laser pulses with particle beams and plasma targets allows for a number of relevant experimental configurations, there are three characteristic setups that are discussed in the literature: laser-particle beam, laser-plasma, and laser-laser.\cite{zhang.pop.2020,Gonoskov_2022} Each of these setups is characterized by a unique interaction geometry and an interplay of SFQED effects with single-particle and collective dynamics. The main difference comes from the role of the laser. While in a laser-particle beam setup, the laser pulse serves as a target for the particle beam (which is in the spirit of studies at high-energy particle colliders), in the laser-plasma and laser-laser setups, the laser field serves not only as a target but also as an accelerator for charged particles. \CORR{Efficient acceleration can be achieved in very different regimes and geometries,\cite{Gonoskov_2022} e.g. in a wake created in laser-plasma interaction or directly by an electric field at the magnetic nodes of a standing wave formed in laser-laser head-on collision. The latter configuration is central in our work.} In the laser-particle beam setup, there are usually no plasma effects at play, since the particle densities in a beam do not reach high values. In contrast, the laser-plasma and laser-laser setups can be tailored to study collective effects. Up to now, only the laser-particle beam configurations was explored experimentally for moderate laser intensities,\footnote{Here, we refer to the laser intensities being moderate as the ones $\sim10^{20}$ W/cm$^2$, whereas high intensities are usually $\sim10^{22-24}$ W/cm$^2$ as envisioned for the next generation multi-petawatt laser facilities.} which means that mostly the photon production and electron beam depletion were detected, with occasional electron-positron pair production in \CORR{conventional accelerator-based} experiments.\cite{bula.prl.1996,burke.prl.1997,poder.prx.2018,cole.prx.2018,mirzaie.natphot.2024,matheron.arxiv.2024, arrowsmith2024laboratory} \CORR{New experiments within this paradigm are the E320 at SLAC\cite{salgado2021single} and the LUXE at DESY\cite{luxe2024technical}, which will explore previously unaccessible regimes of SFQED.}

\begin{figure}
    \centering
    \includegraphics[width=\linewidth]{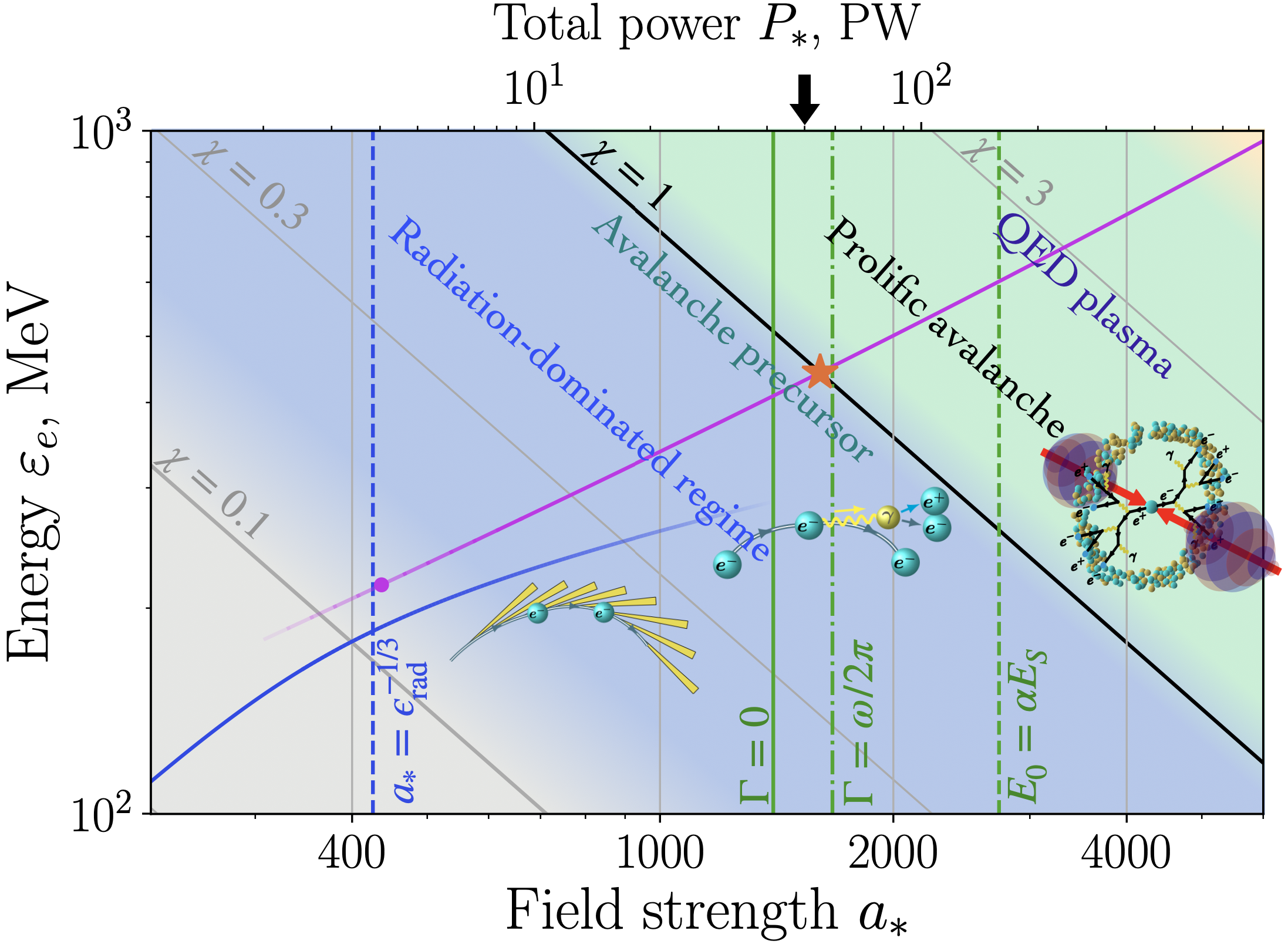}
    \caption{Landscape of the interaction regimes for electrons injected in an electric antinode of a standing wave formed by two counterpropagating multi-petawatt laser pulses. The parameter space is spanned by the total field amplitude (in dimensionless units, $a_*=2a_0$) and the energy $\varepsilon_e=mc^2\gamma_e$ that an electron can gain when accelerated by the electric field of the standing wave. The top horizontal axis shows the corresponding values of the total laser power delivered by two circularly-polarized laser beams focused to the diffraction limit ($f/1$). The arrow at $P_*=50$ PW indicates the planned capabilities of NSF OPAL. The solid curves show $\varepsilon_e$ accounting for radiation reaction within the classical LL model (blue) and the SFQED prediction (magenta), given by Eqs.~\eqref{energy_cl_LL} and \eqref{tem_gem_cem}, respectively. 
    The vertical lines highlight the transition to various interaction regimes. \CORR{Here, the parameter $\erad=2 r_e\omega/3c$ determines the magnitude of the classical RR force [see also Eq.~\eqref{energy_cl_LL}], and $\Gamma$ denotes the avalanche-type cascade growth rate ($N\sim\exp[\Gamma t]$) depending on $\atot$, see also Eq.~\eqref{growth_rate_CP}.} The star-shaped marker corresponds to the point beyond which prolific QED avalanches become accessible. \CORR{The $\chi$ lines are given by $\chi_e=(1/2) a_*\gamma_e \hbar\omega/mc^2$, where $\omega$ corresponds to the laser wavelength, i.e., 0.91 $\mu$m. Under this choice, the line $\chi_e=1$ matches $\cem(a_*)= 1$ at the point $\gamma_{e}\approx\gem(a_*)$, where $\cem$ and $\gem$ are given in Eq. \eqref{tem_gem_cem}.}}
    \label{fig:landscape}
\end{figure}

Since SFQED processes such as photon emission and pair production depend strongly on the EM field strength, it was repeatedly shown in theoretical and computer simulation studies (see the reviews\cite{Di_Piazza_2012,Gonoskov_2022,fedotov2023high,popruzhenko2023dynamics} and references cited therein) that the utilization of high-power laser facilities \cite{danson.hplse.2019} will pave the way towards new unexplored regimes of particle interactions with strong EM fields. Novel multi-petawatt (PW) laser facilities such as ELI-NP,\cite{radier202210, tanaka_mre2020} ELI Beamlines,\cite{Spinka17, weber_mre2017} CoReLS,\cite{sung20174} Apollon,\cite{yao2025characterization} SULF,\cite{li2018339} and the future NSF OPAL \cite{nsf_opal} can be used for efficient generation of ultra-bright high-energy radiation and particle beams by taking advantage of multiple photon emissions, prolific pair production, and re-acceleration of charged particles that have lost energy due to radiation emission (see e.g., Refs. \cite{gonoskov2017ultrabright,magnusson.prl.2019}).

These interaction regimes can be revealed in QED cascades, which are arguably one of the most fascinating SFQED phenomena. A QED cascade is the process of an electron, positron, or photon energy transformation during their interaction with a strong EM field into multiple secondary particles (see Refs. \cite{Di_Piazza_2012,Gonoskov_2022,fedotov2023high,popruzhenko2023dynamics} and references cited therein) and usually consists of a series of nonlinear Compton and Breit-Wheeler events following one after another in quick succession. The two basic cascade types are identified by the interaction mechanism: the \emph{shower} and the \emph{avalanche} (or self-sustained).\cite{mironov2014collapse}

A shower is a sequence of SFQED processes triggered by high-energy charged particles or photons that enter a strong field region.\cite{bulanov2013electromagnetic,blackburn2017scaling,mercuri2021impact,pouyez2025kinetic} In such events, the strong field induces elementary processes, however, its influence on the particle motion is usually negligible. Thus, the initial particle energy is transformed into secondary particles and the total energy of all particles, initial and secondary, remains roughly constant during the interaction. In QED avalanches, charged particles (both initial and secondary) are continuously re-accelerated by the field.\cite{kirk2008,kirk2009,fedotov2010limitations,elkina2011qed} As a result, the total particle number and energy can grow exponentially over time, as the energy is drawn from the field. Both cascades operate in the so-called \emph{radiation-dominated (RD) regime} of interaction. In this regime, the charged particle dynamics in a strong field are dominated by the emitted radiation, and a significant fraction of the field energy can be transferred into high-energy radiation.\cite{bulanov2004interaction,di2010quantum,bulanov.nima.2011,sokolov2011numerical,ji2014radiation,liseykina2025probing,gonoskov2017ultrabright} 

It is rather straightforward to relate the cascade types with the corresponding experimental setups: laser-particle beam interaction is suitable for showers,\cite{pouyez2024multiplicity} whereas avalanches can be generated in laser-laser interaction.\cite{bashmakov2014effect,gelfer2015optimized,grismayer2017seeded,mironov2021} It is worth noting that these associations are not set in stone and different types of cascades can occur either almost simultaneously or concurrently in different interaction geometries (e.g., see Refs. \cite{mironov2014collapse,magnusson2019multiple,samsonov2019laser,samsonov2021hydrodynamical}). 

The above-mentioned multi-PW facilities will allow experimental studies of matter coupled to extreme fields deeply in the classical and quantum RD regimes. 
While abundant production of electron-positron pairs can still be challenging, it seems possible to investigate the transition from the single-particle motion dominated by strong quantum recoil to \textit{avalanche precursors},\cite{magnusson2019multiple} marked by generation of a measurable amount of $e^-e^+$ pairs during the avalanche onset stage. We illustrate the accessible interaction regimes in Fig.~\ref{fig:landscape} and discuss them in more detail in the following section.

Experiment-wise, the most limiting factors are brought by the requirements to reach the highest intensity possible, and the necessity to synchronize at least two counterpropagating laser pulses at the interaction point. While simulations predict that cascades can be potentially triggered at 10 PW scale laser facilities in an ideal scenario,\cite{gelfer2015optimized,jirka2016} the technical constraints on the beam alignment and tight focusing can raise the peak power requirements. Namely, counterpropagating beams have to be tilted to avoid backpropagation of laser pulses, and hard focusing to the diffraction limit $f/1$ can be challenging for a multi-PW laser beam.

In what follows, we investigate the capabilities of the near-future multi-PW facilities for studying particle dynamics deeply in the RD regime and generation of avalanches within the laser-laser interaction scheme. We focus on the anticipated range of NSF OPAL. The preliminary design of this facility suggests bringing two synchronized, tightly focused laser pulses up to 25 PW each to a single interaction point, as required for triggering QED avalanches. We study the interaction of two laser pulses incident on a static target in a realistic configuration, investigate the particle dynamics in the RD regime and the cascade evolution, and identify the final distribution key for designing future experiments.

We rely on full-3D particle-in-cell (PIC) simulations with the code SMILEI\cite{derouillat2018smilei} and perform simulations with tightly focused short laser pulses with realistically achievable parameters. We suggest practical targets for injecting electrons into the laser field focus during the interaction. This can be achieved with a noble gas,\cite{artemenko2017ionization} dense plasma,\cite{grismayer2016} a solid target,\cite{ridgers2012dense, ridgers2013dense, jirka2016} or a high-energy particle-beam.\cite{mironov2014collapse} In this work, we test the first three options and compare the efficiency of each target for the generation of QED avalanches. Finally, we perform a parametric study on laser parameters and investigate the effect of realistic laser beam alignment on the interaction outcomes. This includes tilting of the optical axes to avoid backpropagation of light in the optical systems and non-ideal mutual focusing of two beams on the target.

The paper is organized as follows. We start with a brief review of the physical interaction regimes accessible with ultra-high-intensity lasers in Section~\ref{sec:ii}: the RD regime in the classical (Section~\ref{sec:iia}) and quantum (Section~\ref{sec:iib}) pictures, the basic theory of QED avalanches (Section~\ref{sec:iic}), and the NSF OPAL potential for studying the related effects. In Section~\ref{sec:iii}, we discuss in detail the dynamics of QED cascades triggered by laser pulses with maximal parameters anticipated at NSF OPAL, interacting with targets of three types: a noble gas, a plasma, and a foil. In particular, we first describe the suggested targets (Section~\ref{sec:iiia}), then cover the simulation setup (Section~\ref{sec:iiib}), discuss the target efficiency for electron injection into the laser focus (Section~\ref{sec:iiic}), and compare them in terms of the particle and energy yield in a cascade (Section~\ref{sec:iiid}). In Section~\ref{sec:iiie}, we describe final particle distributions for one target type (krypton gas). In Section~\ref{sec:iv}, we perform a parametric study of the total photon and positron yield with respect to the laser total power, polarization, and focusing below and above the maximum NSF OPAL capabilities. The effect of non-ideal alignment and synchronization of the laser beams is studied in Section~\ref{sec:v}. We conclude our work in Section~\ref{sec:vi} with a summary and outlook.

\section{Physical regimes}
\label{sec:ii}
The dynamics of charged particles in an EM field strongly depend on the field strength and its geometry. A strongly accelerated charge can emit a large amount of radiation, causing radiation reaction (RR), which modifies the particle's motion. In very strong fields, quantum effects in radiation become sizable and QED cascades can be triggered. All these phenomena allow for new interaction regimes. A suitable field configurations for exploring cascades is a standing EM wave, which is a reference case.\cite{kirk2008,fedotov2010limitations,nerush2011laser,grismayer2016,mercuri2025growth} Below, we qualitatively describe the interaction regimes of an electron with such a field at high intensities.

\subsection{Radiation-dominated regime in the classical picture}

\label{sec:iia}
Let us consider an electron in a strong EM standing wave formed by two counterpropagating laser beams. For simplicity, in this section, we assume that the two laser beams have circular polarization (CP) in such a way that the magnetic field $\vec{B}$ vanishes and the electric field vector $\vec{E}$ rotates at the $E$ antinodes. The rotating $\vec{E}$ field admits the solution of classical equations of motion for an electron injected in the center of an electric antinode,\cite{bulanov2011lorentz} which is useful for classifying the interaction regimes.

\CORR{Let us denote the dimensionless amplitude of a rotating $\vec{E}$ field by $\atot=2a_0$, where} 
\begin{equation}
	\label{a0}
	a_0=\frac{eE_0}{mc\omega}.
\end{equation}
Here, $e>0$ is the elementary charge, $m$ is the electron mass, and $E_0$ and $\omega$ are the amplitude and frequency of a single laser beam.
\CORR{Consider an electron injected into the field, initially at rest. According to the classical equations of motion, the electron's Lorentz factor will oscillate with time as $\gamma_e= \sqrt{1+\atot^2\sin^2(\omega t/2)}$.  At $\atot>1$, the electron will become} relativistic within a subcycle of the field. As $\atot$ is increased further, the interaction changes qualitatively. Let us briefly discuss these interaction regimes, which we also illustrate in Fig.~\ref{fig:landscape}. 

Accelerated charges radiate, and at $\atot\gg 1$, the electron motion is modified by RR.\cite{Di_Piazza_2012,Gonoskov_2022,fedotov2023high,popruzhenko2023dynamics} This modification competes with acceleration and limits the maximum energy that the electron can acquire. RR starts to dominate the electron's motion at $\atot\gtrsim \erad^{-1/3}$,\cite{bulanov2011lorentz} when the radiated energy becomes comparable to the electron's energy gain from acceleration. This characterizes the RD regime. Here, $\erad=2 r_e\omega/3c$ is the classical dimensionless parameter determining the magnitude of the RR force, and $r_e=e^2/mc^2$ is the classical electron radius. Precisely, in a rotating electric field, the energy balance condition reads:
\begin{equation}
	\label{energy_cl_LL}
	\atot^2=(\gamma_e^2-1)(1+\erad^2\gamma_e^6),
\end{equation}
which is plotted in Fig.~\ref{fig:landscape} (see the continuous blue line). Note that the curve changes slope at $\atot\sim\erad^{-1/3}$, which marks the transition to the RD regime.

\subsection{Quantum radiation}
\label{sec:iib}
In the setup at hand, the electron energy is essentially determined by $\atot$. As $\atot$ increases, the classical picture of RR breaks down. From the perspective of QED, radiation is a stochastic process with discrete photon emissions causing recoil. This is described as nonlinear Compton emission, $e^\pm+n\gamma_L\rightarrow e^\pm+\gamma$, in which electrons exchange $n\gg 1$ soft photons $\gamma_L$ with the background laser field. Furthermore, in a strong field, the emitted hard photons can create $e^-e^+$ pairs via the nonlinear Breit-Wheeler process, $\gamma+n'\gamma_L\rightarrow e^-+e^+$.

The total probability rates $\WCS$ and $\WBW$ for these elementary processes are controlled by the energy $mc^2\gamma_{e,\gamma}$ and the quantum dynamical parameter $\chi_{e,\gamma}$ of the incident $e^\pm$ or photon, respectively.\cite{ritus1985, baier1968,Di_Piazza_2012,Gonoskov_2022,fedotov2023high} The latter is defined as
\begin{equation}
	\label{chi}
	\chi_{e,\gamma} = \dfrac{1}{E_S }\,\sqrt{ \left(\gamma_{e,\gamma}  \vec{E} + c \vec{u}_{e,\gamma} \times  \vec{B} \right)^2 - \left(\vec{u}_{e,\gamma}\cdot \vec{E}\right)^2 },
\end{equation} 
where $\vec{u}_{e,\gamma}=\vec{p}_{e,\gamma}/(mc)$, $\vec{p}_{e,\gamma}$ are the electron and photon momenta, and $\vec{E}$ and $\vec{B}$ are the local electric and magnetic field components of the laser field, respectively. For $e^\pm$, $\chi_e$ corresponds to the rest frame field strength in units of $E_S$. Here, $E_S=m^2c^3/e\hbar$ is the Sauter-Schwinger field. The amplitude $E_S$ corresponds to an intensity of about $4.6\times 10^{29}\;\text{W/cm$^2$}$.\cite{Di_Piazza_2012,Gonoskov_2022,fedotov2023high} We present the full expressions for $\WCS$ and $\WBW$ in Appendix~\ref{app:rates}.

Within QED, it is natural to characterize the interaction regimes by the value of $\chi_e$. Emission becomes essentially quantum at $\chi_e\gtrsim 0.1$,\footnote{The radiation intensity calculated within the classical and strong-field QED approaches overlaps at lower values of $\chi_e$.\cite{ritus1985, baier1968,Di_Piazza_2012,Gonoskov_2022,fedotov2023high, bulanov2013electromagnetic, niel2018}} at which point the particle dynamics change. If the field is well below the Sauter-Schwinger limit ($E\ll E_S$) and the electron is ultrarelativistic, the electron motion can still be viewed as semiclassical. Namely, particles propagate along classical trajectories,\cite{baier1968} but the RR is accounted for in random photon emission events\cite{elkina2011qed}. These events can be treated as point-like within the locally constant field approximation,\cite{Di_Piazza_2012,Gonoskov_2022,fedotov2023high} which we assume valid for the interaction regimes discussed in this work, as it generally requires $a_0\gg 1$ (see also Appendix~\ref{app:approximations} for a more detailed discussion of the approximations in use).

During the electron acceleration, $\chi_e$ varies and can increase with time, as well as $\gamma_e$. Hence, the emission probability rate is parametrized as $\WCS(t)=\WCS[\gamma_e(t), \chi_e(t)]$, and, more importantly, it can continuously grow along an electron's semiclassical trajectory. As a result, the field energy can be efficiently converted into radiation. This is a striking feature of the standing-wave field configuration that defines the interaction. Notably, such restoration of $\chi_e$ is not revealed, for example, in the interaction of an electron with a single plane wave, as $\chi_e$ is conserved during semiclassic motion in between quantum emissions.\cite{landau2013classical} The quantum parameter of each particle produced in an elementary QED event (either photon emission or pair production) is smaller than that of the particle initiating the event. These two points combined make the plane-wave setup (namely, a \CORR{weakly or moderately focused single laser}), in principle, unsuited to access the important regimes investigated here. \CORR{Notably, at very strong focusing, $\chi_e$ can grow for electrons injected in the focus.\cite{mironov2021} Furthermore, interaction with a dense plasma can change the dynamics significantly, opening access to different interaction regimes, including avalanche-type cascades.\cite{samsonov2019laser,samsonov2021hydrodynamical}}

Taking this into consideration, we can estimate the time scale $\tem$ at which the accelerated electron emits as $\int^{\tem} \WCS(t) dt=1$. By $\tem$, the electron acquires $\gem=\gamma_e(\tem)$ and $\cem=\chi_e(\tem)$. As a result, the properties of emission and the interaction regime are defined by these quantities; $\cem$ can be small or large compared to unity, depending on $\atot$ and other field parameters. In particular, in a CP field, we can approximate the field dependence at $\cem< 1$ as:\cite{mercuri2025growth}
\begin{equation}
	\label{tem_gem_cem}
    \begin{split}
	\omega\tem &\simeq 1.7\times\frac{\sqrt{\omega\tau_C  }}{\alpha} \sqrt{\dfrac{\alpha \aS}{\atot}},\\
	\gem &\simeq 1.7\times\sqrt{\dfrac{\atot}{\omega \tau_C \alpha \aS}},\quad \cem\simeq 1.4\times\dfrac{\atot }{\alpha \aS},
    \end{split}
\end{equation}
where $\alpha$ is the fine-structure constant, $\tau_C=\hbar/mc^2\approx1.29\times10^{-21}$ sec is the Compton time, and we introduced $\aS=eE_S/mc\omega\equiv 1/\omega\tau_C$. The approximate identities in Eq. \eqref{tem_gem_cem} are valid at $\omega\tem<1$. As $\tem$ decreases with $\atot$, they are robust at strong fields such that $\atot\gg 1$. We plot the full expression of $\gem$, i.e., beyond the asymptotic expressions in Eq.~\eqref{tem_gem_cem} (see also Refs.~\cite{mercuri2025growth,mironov2021}) as a function of $\atot$ in Fig.~\ref{fig:landscape} (see the magenta line; the round point denotes the validity limit $\omega\tem=1$ from below). In the crossover region $\cem\gtrsim 0.1$, $\gem$ matches the classical expression in Eq. \eqref{energy_cl_LL} within an order of magnitude. However, the classical theory overestimates the emission losses as $\atot$ is increased. This is a well-known feature seen in classical synchrotron radiation, see, e.g. Ref.\cite{ritus1985} We also note that the validity of the classical description can be extended by introducing a quantum correction factor.\cite{erber1966high, ridgers2014modelling, niel2018}

\subsection{QED cascades}
\label{sec:iic}
When $\cem$ approaches 1 (marked by a star in Fig.~\ref{fig:landscape}), the probability of pair creation by the emitted photons becomes high. This signifies the transition from single-electron dynamics in the RD regime to the onset of QED avalanche-type cascades, which is essentially a many-body regime. As the initial and secondary charges continuously experience acceleration and emit hard photons, $e^-e^+$ pair yield can reach a steady exponential growth in time, $N\sim \exp(\Gamma t)$,\cite{kirk2008, kirk2009, fedotov2010limitations, mercuri2025growth} where $\Gamma$ is the so-called cascade growth rate. Estimates show that the field scale for reaching high-multiplicity cascades is of the order $\atot\sim \alpha \aS$\cite{fedotov2010limitations,elkina2011qed}  (shown in Fig.~\ref{fig:landscape}). However, it was demonstrated with simulations\cite{ridgers2012dense, ridgers2013dense, mironov2014collapse, gelfer2015optimized, grismayer2017seeded, artemenko2017ionization, yu2018qed} and recent theoretical analysis\cite{mironov2021, mercuri2025growth} that avalanche-type cascades can onset at significantly lower fields.

The cascade dynamics can be split into two stages: onset and a (quasi-)steady state. The former is highly nonstationary, as the characteristic quantities (such as the average energy and $\chi_e$) in the cascade change rapidly. If the field duration is long enough, the variation of these characteristics saturates, and the cascade reaches the second stage, in which the time dependence of $\WCSBW$, $\Gamma$, and other quantities vanishes (or they vary adiabatically at a ``slow'' time scale determined by the field envelope).

The spatial distribution of the field, e.g., if laser beams are tightly focused, also affects the cascade growth rate $\Gamma$.\cite{Tamburini_2017} As the interaction area for such fields is small, $e^\pm$-s and $\gamma$-s migrate from the focal area during the interaction at rates $\nu_{e,\gamma}\propto \omega$.\cite{jirka2016} This limits the pair yield in a cascade at lower fields,\cite{tang2024finite} and imposes a threshold condition for the steady state formation.\cite{mercuri2025growth} Overall, $\Gamma$ depends on the field parameters parametrically via the rates of particle production and migration:\cite{mercuri2025growth}
\begin{equation}
	\label{growth_rate_general}
	\begin{split}
		\Gamma &= \dfrac{\WBW + \nu_e + \nu_\gamma}{2} \\
		&\,\,\,\times \left[-1+\sqrt{1 + 4\frac{\WBW(2\WCS - \nu_e) -\nu_e\nu_\gamma}{(\WBW + \nu_e + \nu_\gamma)^2}} \right],
	\end{split}
\end{equation}
where $\WCSBW=\WCSBW(\gem,\cem)$ are the effective process rates in a cascade.\footnote{This approximation for the $\WCSBW$ in a cascade is valid for fields $E\lesssim\alpha E_S$, and at higher fields should be modified, see Ref.\cite{mercuri2025growth} for details. However, such high fields are out of the scope of this work.} The threshold condition then reads $\Gamma=0$, which can be viewed as an implicit equation for the field strength. In particular, for the CP Gaussian beams focused to the diffraction limit ($f/1$), $\Gamma$ near this threshold can be written explicitly:
\begin{equation}
	\label{growth_rate_CP}
	\Gamma\simeq \frac{\omega}{\pi}\left[ -1 + \frac{1.52\pi^2 \alpha \atot}{\omega \tau_C \aS}\exp\left(-1.92\frac{\alpha \aS}{\atot}\right)\right],
\end{equation}
where we estimated the electron and photon migration as $\nu_{e}\sim 2c/w_0+\omega/(2\pi)$ and $\nu_\gamma\sim2c/w_0$ with the beam waist $w_0\sim2\pi c/\omega$. In Fig.~\ref{fig:landscape}, we plot the field strength $\ath$, corresponding to the threshold condition $\Gamma(\ath)=0$ (vertical solid green line). At $f/1$, $\ath\approx 1400$, which corresponds to $P_*\approx 42$ PW.

We emphasise that for efficient particle production, the cascade should reach a steady state. In short laser pulses, this is not guaranteed, especially if the field strength is close to $\ath$. Still, $e^-e^+$ pair yield can be relatively high already at the onset stage. We denote this subregime as the \textit{avalanche precursor}: a transition from the quantum RD regime to high-multiplicity avalanches. For the sake of an estimate, one may assume that it extends up to the field strength $\atot$ such that $\Gamma(\atot)=\omega/2\pi$ (shown in Fig.~\ref{fig:landscape} as the vertical green dash-dotted line). This condition guarantees a high particle yield in one field cycle.

Finally, at very strong fields, $\atot>\alpha \aS$, the particle production rate becomes so high, that the created $e^-e^+$ pairs can rapidly form a dense relativistic plasma and even potentially screen the laser field.\cite{nerush2011laser, grismayer2016}

\subsection{NSF OPAL potential}
\label{sec:iid}
\begin{table}
	\caption{Expected parameters of a single Alpha beam at NSF OPAL. The possibility of two such beams at the interaction point is included in the facility design. The value of $a_0$ is calculated for the wavelength $\lambda=0.91$ $\mu$m.}
	\label{tab:nsf_opal_params}
	\begin{tabular}{l| c}
		\hline\hline
		Power $P_0$ & 25 PW\\
		Intensity & $5\times 10^{23}$ W/cm$^2$\\
		Energy & 500 J\\
		Pulse duration & 20 fs (FWHM)\\
		Focal spot & 2.3 $\mu$m (f/2, FWHM) \\
		Polarization & Linear or Circular\\
		Peak $a_0$ at f/2 LP & 550 \\
		Peak $a_0$ at f/2 CP & 389 \\
		Pulse co-timing & 10 fs (rms)\\
		Co-pointing precision &	1.5 $\mu$m\\
		\hline\hline
	\end{tabular}
\end{table}
The anticipated capabilities of NSF OPAL \cite{nsf_web} will allow for the exploration of most of the regimes charted in Fig.~\ref{fig:landscape}. The relevant parameters are summarized in Table~\ref{tab:nsf_opal_params}. The maximum total power $\Ptot=2P_0=50$ PW delivered by two laser pulses, called Alpha beams (each of power $P_0$) is marked by an arrow on the top horizontal axis of Fig.~\ref{fig:landscape}. Based on the estimates discussed above, the electron-laser interaction can be studied deep in the RD regime, as well as the quantum RR regime. At the peak potential, seed electrons can gain $\cem\gtrsim 1$, therefore, the avalanche precursors are within reach.

Pair production in cascades is sensitive to the peak field strength $\atot$. At fixed peak power, the maximum $\atot$ is higher (by a factor of $\sqrt{2}$) for a linearly polarized (LP) wave as compared to CP. The classical electron dynamics in LP standing waves is known to be chaotic\cite{esirkepov2014} and qualitatively changes with growing $\atot$,\cite{gonoskov2014anomalous} which complicates analytic considerations.  Still, qualitatively, the regimes of electron-laser interaction and their characteristic scales are similar to the CP case. In particular, electrons can be accelerated efficiently, so that $\chi_e$ can grow rapidly. Therefore, the RD interaction regime\cite{magnusson2019multiple} and avalanche-type cascades\cite{bashmakov2014effect,grismayer2017seeded} are also accessible. For a more general analysis of field configurations suitable for triggering cascades see Refs.\cite{mironov2021,mercuri2025growth}

\section{Generation of avalanche-type cascades}
\label{sec:iii}
\begin{figure}
	\centering
	\includegraphics[width=\linewidth]{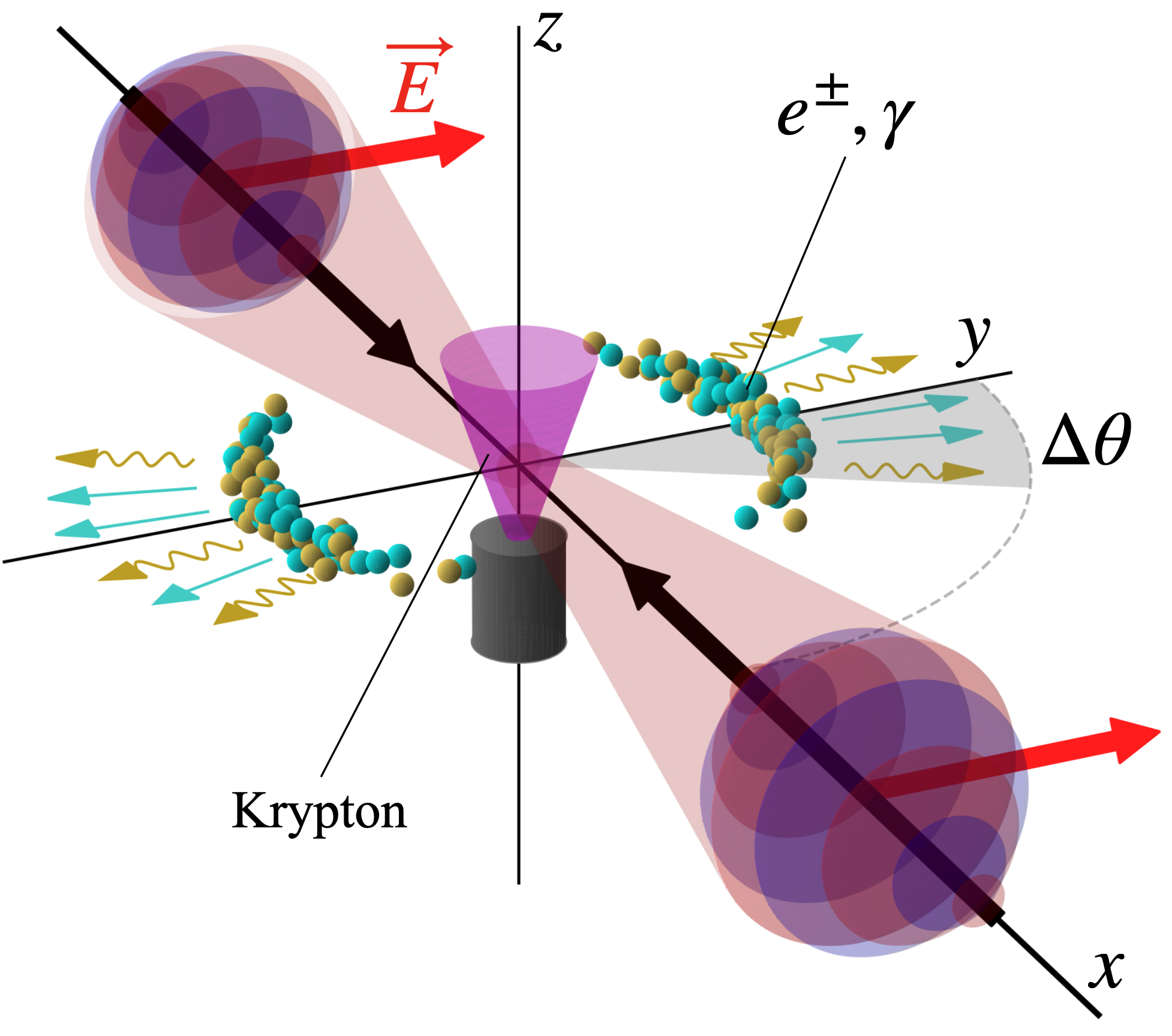}
	\caption{\CORR{Scheme of the interaction: two incident LP laser pulses (with the parameters listed in Tab.~\ref{tab:nsf_opal_params}) form a standing wave and ionize a krypton gas. Electrons are injected into the laser focus and trigger an avalanche-type cascade. The generated $e^\pm$ and $\gamma$-photons outgoing from the interaction point are depicted with blue and yellow spheres, respectively. $\Delta\theta$ denotes a polar angle opening (see Sec.~\ref{sec:iiie}).}}
	\label{fig:scheme}
\end{figure}

We first study the possibility of generating QED cascades with two LP\footnote{Throughout, in our simulations, we consider only linear polarization, except for Section~\ref{sec:iv}, where we explicitly compare the results for the LP and CP fields.} laser pulses at the maximal parameters anticipated at NSF OPAL (see Tab.~\ref{tab:nsf_opal_params}). In this section, we consider the best-case scenario laser field configuration. Namely, the laser pulses counterpropagate along the same axis and form a standing wave, their polarization axes match, the focal points coincide at $\vec{r}=\vec{0}$, and the pulses are perfectly synchronized in time. This guarantees that the electric field components add up constructively and reach a maximum in the focal center. The effect of the tilted optical axes, misalignment, and pulse mistiming is discussed in Sec.~\ref{sec:iv}.

In this field configuration, we test three realistic options for injecting seed electrons into the focal center: via ionization of a krypton gas, a plasma slab, and a thin aluminium foil. \CORR{The interaction scheme is illustrated in Fig.~\ref{fig:scheme} for the former case. For each target,} we identify the development of a cascade with high photon and positron yield. We briefly discuss the temporal evolution of the cascade parameters and compare the target efficiency in terms of the positron yield.

The analysis is based on 3D PIC simulations with the code SMILEI,\cite{derouillat2018smilei} which includes Monte-Carlo modules for the strong-field QED effects and strong-field ionization. This approach allows us to describe the spacetime evolution of the strongly-focused laser field in accordance with Maxwell's equations and treat the plasma dynamics in the RD regime consistently.

\subsection{Targets}
\label{sec:iiia}
\begin{figure}
	\centering
	\includegraphics[width=\linewidth]{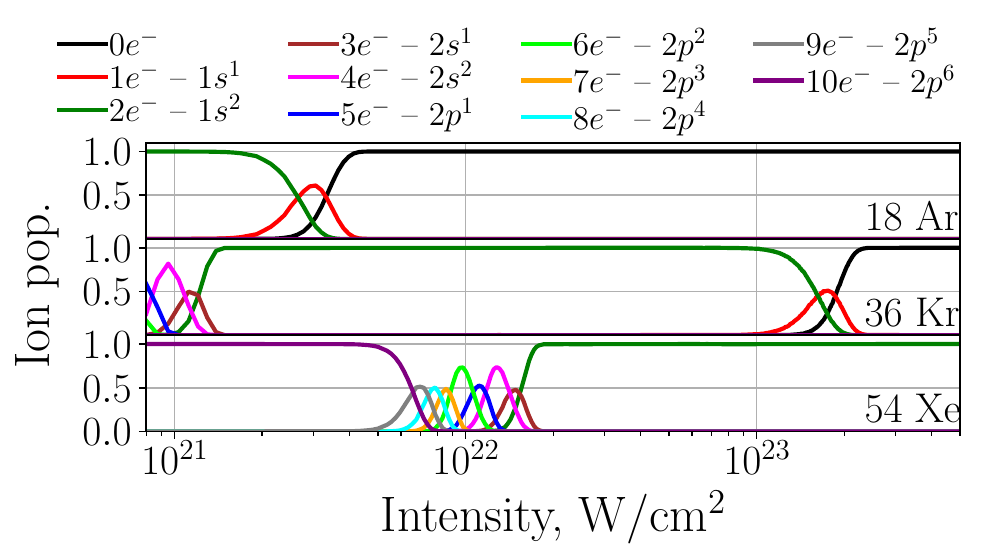}
	\caption{Dependence of ion populations of noble gases on the laser intensity as a result of tunneling ionization. For a laser interacting with the gas, these curves map the field intensity profile to the expected densities of resulting ions.}
	\label{fig:atoms}
\end{figure}

\begin{table*}[]
	\centering
	\caption{Comparison of the parameters for three different targets, and the corresponding positron yield and energy transfer in avalanche-type cascades. The krypton and plasma targets are initially homogeneously distributed in a ball of radius $r=0.75w_0\approx1.5\lambda$. The aluminium foil is shaped as a disk of thickness $d=0.2w_0$ along the laser propagation axis, and of radius $r=0.75w_0$ in the transverse direction. $\Nsf$ is the number of seed electrons in the focus calculated at time $t=-0.5T$. Here, we use $\varepsilon_{\rm laser}\approx 1000$ J to denote the total laser energy, $\varepsilon_{\gamma}$ --- the photon energy, and $\varepsilon_{e^\pm, i}$ --- the total energy of electrons, positrons, and ions.}
	\label{tab:seeding}
	\begin{tabular}{lccc}
		\hline\hline
		Target & Kr gas  & Plasma & Al foil \\\hline
		Density $n_0$ & $3.74\times 10^{18}$ cm$^{-3}$ & $10n_c$ & $582n_c$ \\
		Size, $\mu$m & $r=1.34$ & $r=1.34$ & $d=0.356$ \\
		$N_{\rm seed}^{\rm foc}$ & $2.89\times 10^6$ & $6.23\times10^{9}$ & $1.16\times 10^{11}$ \\
		$N_{e^+}$ & $3.71\times 10^6$ & $7.44\times10^{9}$ & $6.88\times 10^{10}$ \\
		$N_{e^+}/N_{\rm seed}^{\rm foc}$ & 1.28 & 1.19 & 0.591 \\
		Energy conversion $\varepsilon_\gamma$, \% of $\varepsilon_{\rm laser}$ & $\sim 10^{-3}$ & 2.06 & 25.97 \\
		Energy conversion $\varepsilon_{e^\pm,i}$, \% of $\varepsilon_{\rm laser}$ & $\sim 10^{-6}$ & 0.27 & 4.41 \\
		\hline\hline
	\end{tabular}
\end{table*}

We test three different targets at the interaction point to inject electrons into the laser focus. First, we consider a heavy-atomic noble gas. It is subject to strong-field (tunneling) ionization, \cite{popov_ufn2004, poprz_jpb14} and the threshold field required for ionizing electronic subshells grows with the ionization potential. When the gas is irradiated by an intense laser field, the outer shells are instantly ionized, and the corresponding electrons are quickly expelled from the interaction area.\cite{Tamburini_2017,sampath2018towards} However, deeper subshells of heavy elements can be ionized only by a strong field. The key idea is to select an atom with inner subshells that can be ionized only in the focus, where the field is close to its maximum. Then, the extracted electrons will be exposed to a strong field and trigger the cascade.\cite{artemenko2017ionization} 

In Fig.~\ref{fig:atoms}, we plot the expected ion populations depending on the local field intensity for three noble gases. \CORR{To obtain these curves, we used the Perelomov-Popov-Terent'ev formula for strong-field ionization derived for small values of the Keldysh parameter \cite{perelomov_jetp1966,perelomov_jetp1967,perelomov_jetp1967b} (see also Ref.\cite{ammosov_jetp1986}). In the version of SMILEI PIC code used here, this formula is implemented following the algorithm suggested in Ref.\cite{nuter_pop2011} Let us note that we do not take into account barrier suppression corrections, which are expected to be important for the dynamics of outer shell ionization by an incident laser pulse.\cite{artemenko2017ionization, ouatu_pre2022, mironov2025strong} As we suggest below, in the parameter range of NSF OPAL, it is beneficial to use the $1s$-subshell of krypton as a source of seed electrons triggering a cascade. Ionization of these atomic states appears insensitive to barrier suppression or other corrections to the strong-field ionization model.\cite{mironov2025strong}}

For ${}_{36}$Kr, the threshold for ionizing the two last $1s$ electrons is below but still close enough to the range of intensities relevant for NSF OPAL. Therefore, in the event of Kr$^{34+}$ ionization, these electrons will be injected directly into the focal area. For these reasons, we choose krypton as a candidate for seeding avalanches in our setup. We set the gas density to $3.74\times 10^{18}$ cm$^{-3}$ in the simulation. Assuming a fully ionized gas (initially neutral), this would correspond to the density of $\approx 0.1n_c$ for electrons, where $n_c=\varepsilon_0 m \omega^2/e^2\approx 1.35\times10^{21}$ cm$^{-3}$ is the critical plasma density. However, note that this value is not reached at the interaction point, as the electron density rapidly varies in time and space during the interaction, see the details below. As we are interested specifically in the interaction dynamics in the strong-field region, to optimize the simulation time, we use a small ball-shaped target of radius $r=0.75w_0\approx 1.34$ $\mu$m. We also consider initially charged ions Kr$^{26+}$, as outer electrons are extracted at earlier stages and do not participate in the interaction in the focal center.\footnote{We performed additional simulations to confirm this. Simulations with an initially neutral target and with a pre-charged target result in the same number of seed electrons occurring in the focal center of the laser field.}  Let us mention that our simulations show that, potentially, argon can also be used for triggering cascades. However, when compared to krypton, the number of seed electrons injected in the focal area will be lower for the studied here setup parameters. Using xenon could be efficient at higher intensities.

Next, we consider plasma seeding, and take as well a small ball-shaped target of radius $r=0.75w_0\approx 1.34$ $\mu$m, and set the plasma density to $10n_c$.\footnote{The possibility of pre-generating such plasma at the interaction point is a nontrivial question, however, out of the scope of this work.} Note that under the considered parameters, the total laser field remains unaffected during the interaction with such a plasma target.

Finally, for the aluminium foil, we assume that the target is a thin disk of thickness $d=0.2w_0\approx 0.356$ $\mu$m and $r=0.75w_0\approx 1.34$ $\mu$m in the transverse direction. The laser pulses are normally incident on the foil surface. The targets' parameters are summarized in Tab.~\ref{tab:seeding}. Note that the choice of the target thickness satisfies the condition $a_0>\pi(n_0/n_c)(d/2)/\lambda\approx350$, ensuring that two pulses can penetrate the target from its sides.\footnote{This condition on the target thickness and density implies that a single laser pulse of strength $a_0$ can separate (i.e. push) all of the electrons from ions \cite{vshivkov.pop.1998,bulanov.pop.2016}.}

\CORR{To conclude the discussion about targets, let us estimate the relevance of QED showers linking bremsstrahlung and Bethe-Heitler processes in the targets. For this, we use a simple estimate of the radiation length\cite{bethe1934stopping} $X_0=1/[4\alpha r_e^2 n_0 Z(Z+1) \log(183 Z^{-1/3})]$  (see also Ref.\cite{pouyez2025towards} for recent progress on the theory of showers in matter). Under normal conditions, one obtains for aluminium ($Z=13$, $n_0\approx1.6\times 10^{23}$ cm$^{-3}$) $X_0\sim 3$ cm and orders magnitude larger for krypton. Hence, by using smaller-sized targets, it is possible to exclude the development of showers generated by the bremsstrahlung and Bethe-Heitler processes, which is achievable in future experiments.}

\subsection{Simulation setup}
\label{sec:iiib}
\begin{figure}
	\centering
	\includegraphics[width=\linewidth]{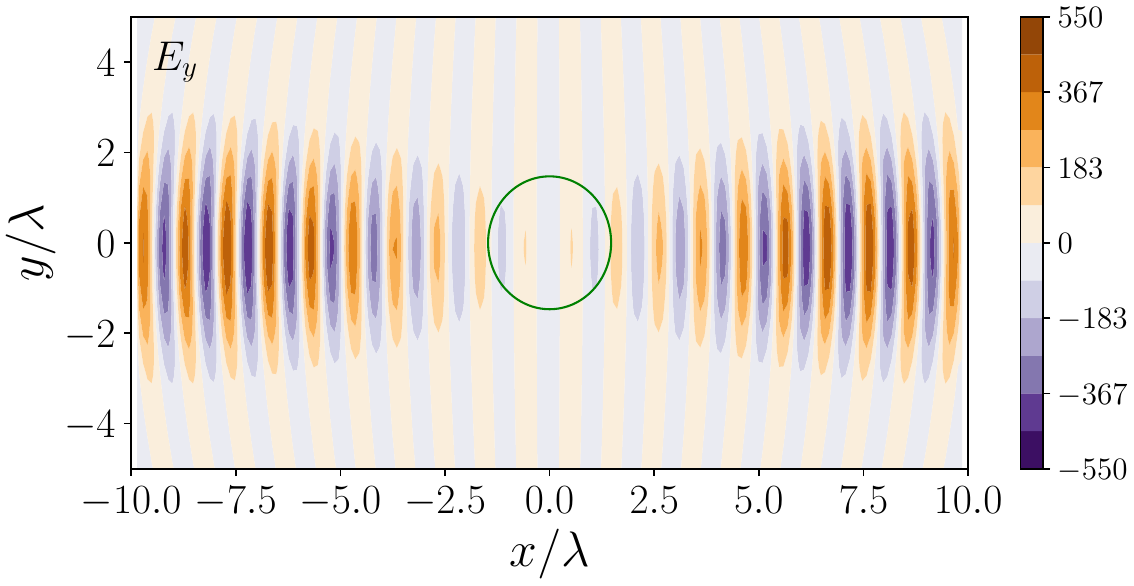}
	\caption{\CORR{Distribution of the $E_y$ component of the field of two counterpropagating laser pulses in a plane $z=0$ before the interaction. The green circle shows the initial shape of the krypton and $10n_c$-plasma targets used in simulation (see also Tab.~\ref{tab:seeding}).}}
	\label{fig:field1}
\end{figure}

\begin{figure*}
	\centering
	\includegraphics[width=0.329\linewidth]{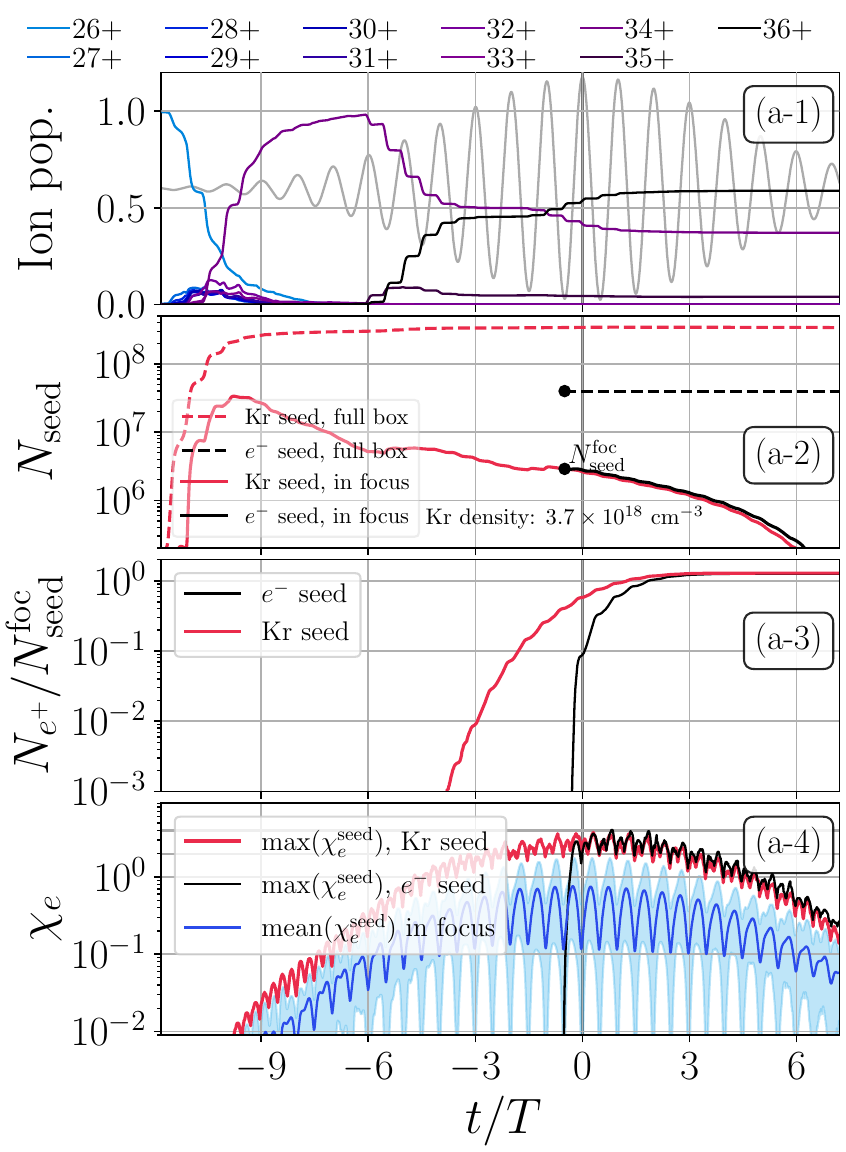}
	\includegraphics[width=0.329\linewidth]{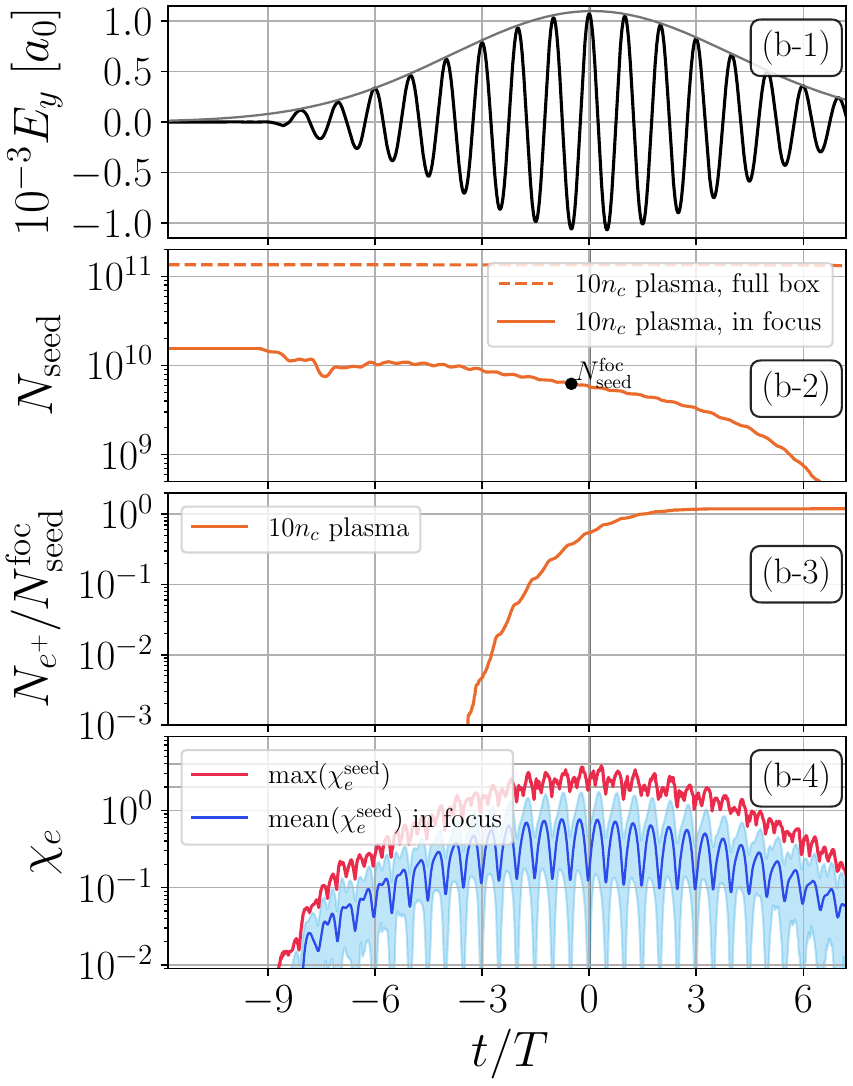}
	\includegraphics[width=0.329\linewidth]{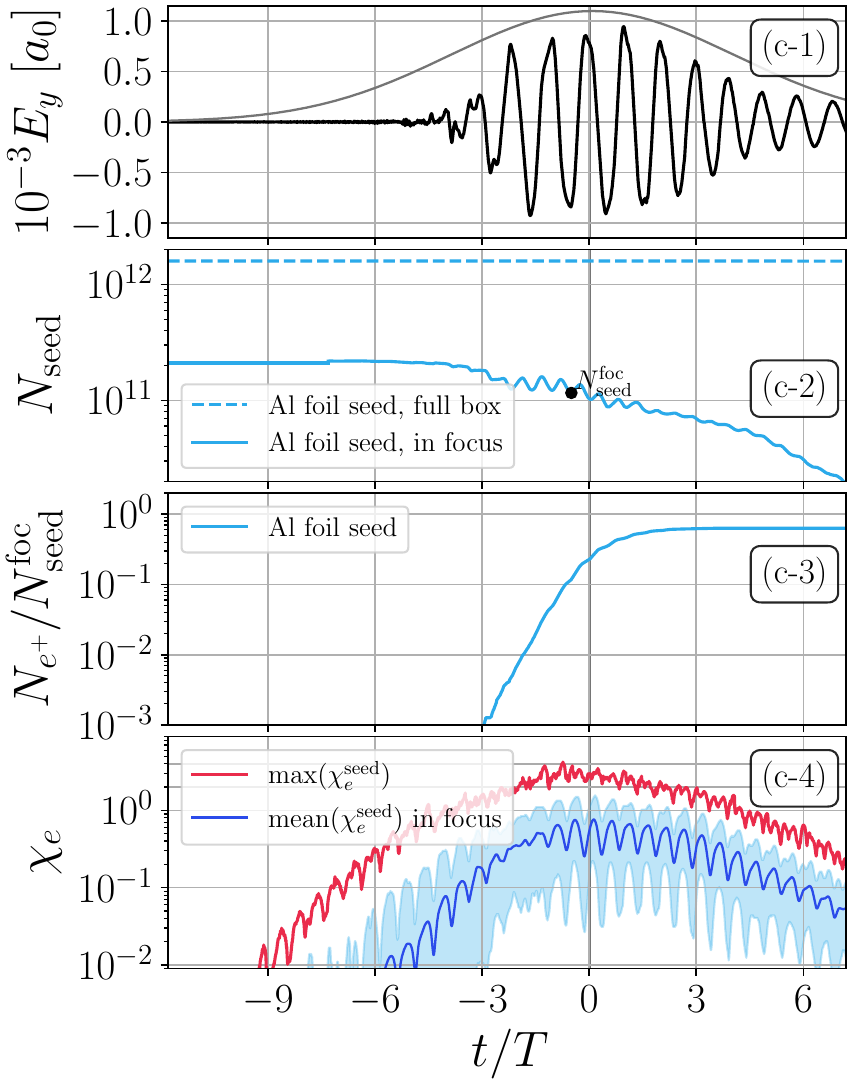}
	\caption{Evolution of the cascade parameters for seeding with: (a) a krypton gas, (b) a plasma of density $10n_c$, (c) an aluminium foil. For each case, the plots show (from top to bottom): 1) field strength at the focal center [in (a-1), the field strength is shown in gray in arb. units. and we also present ion populations], 2) the number of seed electrons in the simulation box (dashed) and in the focus (solid; the focal volume is defined by $-0.5\lambda\leq x,\,y,\,z\leq 0.5\lambda$); the black markers define the number of seed electrons $\Nsf$ in the focus calculated at time $t=-0.5T$, 3) the number of positrons normalized to $\Nsf$, 4) $\chi_e$ of seed electrons in the focal spot (the light-blue band shows the $90$-percentile of the seed electron distribution in $\chi_e$). In (a), we also show the simulation results for an idealized $e^-$ target with black solid lines (``$e^-$ seed'' in the legend), see Section~\ref{sec:iva} for more details. The data is obtained from 3D SMILEI PIC simulations, in which two LP laser pulses counterpropagate, and for each pulse: $a_0=550$, focusing $f/2$ (peak power is $P_0=25$ PW per pulse), duration 20 fs. The target parameters and results for all cases are summarized in Table~\ref{tab:seeding}.}
	\label{fig:seeding}
\end{figure*}

\CORR{We perform 3D PIC simulations with the code SMILEI.\cite{derouillat2018smilei} Let us briefly describe the laser field setup.} Each laser pulse is strongly focused at $\vec{r}=\vec{0}$ and propagates along either the positive or negative $x$-direction. The field wavelength is set to $\lambda=0.91$ $\mu$m throughout and the corresponding period is indicated as $T=\lambda/c$. For each pulse, the temporal envelope is Gaussian with duration $\tau_{\mathrm{FWHM}}=20$ fs (per intensity profile), so that the components of the field are $\propto a_0\exp(-t^2/\tau^2)$, where $\tau=\tau_{\mathrm{FWHM}}/\sqrt{2\ln 2}$. The spatial profile in the focal plane is also Gaussian in the transverse direction with waist $w_{\mathrm{FWHM}}$ (intensity profile), so that the field components are $\propto a_0\exp(-r^2/w_0^2)$ \CORR{in the focal plane $x=0$}, where $w_0=w_{\mathrm{FWHM}}/\sqrt{2\ln 2}$. In this Section, we consider $f/2$ focusing, so $w_{\mathrm{FWHM}}=2.3$ $\mu$m ($w_0\approx 1.96\lambda$). The power is fixed to $P_0=25$ PW per pulse and the polarization is linear along the $y$-direction, so the maximum field amplitude is $a_0=550$ and the corresponding intensity $I_0\approx 5\times10^{23}$ W/cm$^2$ (hence, for the total field, $\Ptot=50$ PW, $\atot=1100$, and intensity $I_{\mathrm{total}}\approx 2\times10^{24}$ W/cm$^2$). Under these parameters, a single laser pulse has an energy of $\approx 500$ J. 

\CORR{The field components are calculated by solving Maxwell's equations numerically with the method suggested in Ref.\cite{thiele2016boundary} We specify the transverse magnetic field components in the focal plane, so that $B_y|_{x=0}=0$ and the $B_z$ component for the pulse propagating in the positive $x$ direction reads (in the dimensionless units):
\begin{equation}
    \label{eq:Bz}
    B_z|_{x=0} =a_0 e^{-(y^2+z^2)/w_0^2} e^{-(t/\tau)^2} \cos t,
\end{equation}
We flip the sign of $B_z|_{x=0}$ for the counterpropagating pulse, so that $x=0$ corresponds to the electric antinode center of the total standing-wave field. Before running a full simulation of the interaction with a target, the field of each laser pulse is propagated backwards from the focal plane to the corresponding simulation box boundary. This provides boundary conditions ensuring the field shape at the focus. Then, all the field components are evaluated in the PIC loop at all time-space points during the simulation run. See also the \texttt{LaserOffset} option description in the SMILEI code manual.\cite{smilei_manual} A snapshot of the numerically calculated $E_y$ component distribution before the interaction is shown in Fig.~\ref{fig:field1}.} 

\CORR{The expected contrast at NSF OPAL is at the level of $>10^{10}$ in the far field up to 100 ps before the main pulse (the facility design is based on optical parametric chirped-pulse amplification technique). We assume that under these parameters, the target (in particular, a foil) survives the interaction with the prepulse, namely, it does not expand significantly before the arrival of the main pulses. Therefore, in this work, we do not simulate the interaction with the pedestal and consider only the main femtosecond-scale pulses. Notably, in the double-pulse geometry, the electrons are compressed by the ponderomotive pressure from the two sides of the target, which also helps to keep the target confined before the main pulse arrival.}

\CORR{We set PIC simulation parameters as follows.} The simulation box is sized $20\lambda$ in the $x$-direction (along the laser pulse propagation) and $40\lambda$ in the $y$- and $z$-direction. The laser field focus and the target are located at the box center at the origin $\vec{r}=\vec{0}$. The time $t=0$ is chosen as the moment when the laser envelope centers reach $\vec{r}=\vec{0}$. At the initial stage of each simulation, the target is not exposed to the laser field. The laser pulses are injected from the box boundaries and are incident on the target from the sides during the run as described above. 

The spatial and temporal resolutions are set to $\mathtt{resx}=\lambda/16$ and $\mathtt{rest}=T/160$, respectively. In simulations with the aluminium foil target, the spatial resolution in the longitudinal $x$-direction is set to $\mathtt{resx}=\lambda/200$ to resolve the corresponding plasma wavelength, and $\mathtt{rest}$ is increased accordingly to satisfy the Courant condition. We use Monte Carlo modules to describe radiation and pair creation (via the nonlinear Compton and Breit-Wheeler processes, respectively), as well as the ionization process. The value of $\mathtt{rest}$ is chosen to resolve the quantum effects. Finally, the number of particles per cell is varied in simulations depending on their computational complexity, but is kept between 8 and 16. We ensured the overall stability of our results under the variation of these parameters.

\subsection{Injection of electrons and the cascade onset}
\label{sec:iiic}

Let us compare the simulation results for the three targets introduced above. In Fig.~\ref{fig:seeding}, we show the temporal evolution of the cascade parameters for the three cases.

Let us first discuss the interaction dynamics with the krypton gas target. In Fig.~\ref{fig:seeding}(a-1), we show the ion population evolution. The curves correspond to the instantaneous ion numbers normalized to the total number of krypton atoms. As the laser field starts interacting with the gas, all states up to Kr$^{34+}$ get rapidly ionized before the peak field arrives at the target. Note that the final krypton ion states are populated by Kr$^{34+}$ and Kr$^{36+}$. As the last two $1s$ electrons require a higher field for ionization, these electrons will appear only in the focal center. Note that the two $1s$ electrons have close potentials and therefore in practice are often ionized together, see also Fig.~\ref{fig:atoms}. The outer volume of the focus, where the field is weaker, are populated by Kr$^{34+}$ and lower charge states.

For brevity, by \textit{seed} we call the electrons that are extracted during the ionization process (for krypton) or are initially contained in the target (for the plasma and foil) before the interaction with the laser field. In SMILEI, it is possible to track their dynamics independently. In Fig.~\ref{fig:seeding}(a-2), one can see that as the ionization process develops, the corresponding number of seed electrons $\Ns$ grows (see the red dashed line) and saturates after the maximal field is reached. 

A cascade can be triggered only by seed electrons located near the focal center. Let us define the focal volume by a box of volume $\lambda^3$ at the center and track the number of electrons in it. This number is shown by the red solid line in Fig.~\ref{fig:seeding}(a-2). After a sharp increase, the number of seed electrons in the focus gradually decreases as they are pushed by the field. The two bumps near $t\approx-4.5T$ and $-0.5T$ correspond to the ionization of the $1s$ electrons in the focus. The maximum and mean $\chi_e$ for electrons in the focus exceed unity [see Fig.~\ref{fig:seeding}(a-4)]. Therefore, such electrons can trigger an avalanche-type cascade.

In Figs.~\ref{fig:seeding}(b) and (c), we show the dynamics of the seed electron number and $\chi_e$ in the focus for the plasma and foil targets, respectively. The overall qualitative behaviour is similar: as the field impinges on the target, the electrons in the focus are pushed by the field, and at the same time the corresponding maximum and mean values of $\chi_e$ grow above unity. The main difference from the krypton target is a higher density of seed electrons. 

It is convenient to introduce the number of seed electrons in the focus $\Nsf$ calculated at a fixed time close to the cascade onset. This allows for a consistent measure of different cascade parameters per seed electrons. We choose $t=-0.5T$, as at this moment the laser field amplitude is close to its maximum, and the conditions are favourable for the cascade development. We explicitly show $\Nsf$ for each target in Figs.~\ref{fig:seeding}(a-2), (b-2), and (c-2) and present the corresponding values in Tab.~\ref{tab:seeding}. Because $\Nsf$ is proportional to the initial target density, a foil provides the highest value.

\subsection{Positron yield and energy transfer}
\label{sec:iiid}
\begin{figure}
	\centering
	\includegraphics[width=\linewidth]{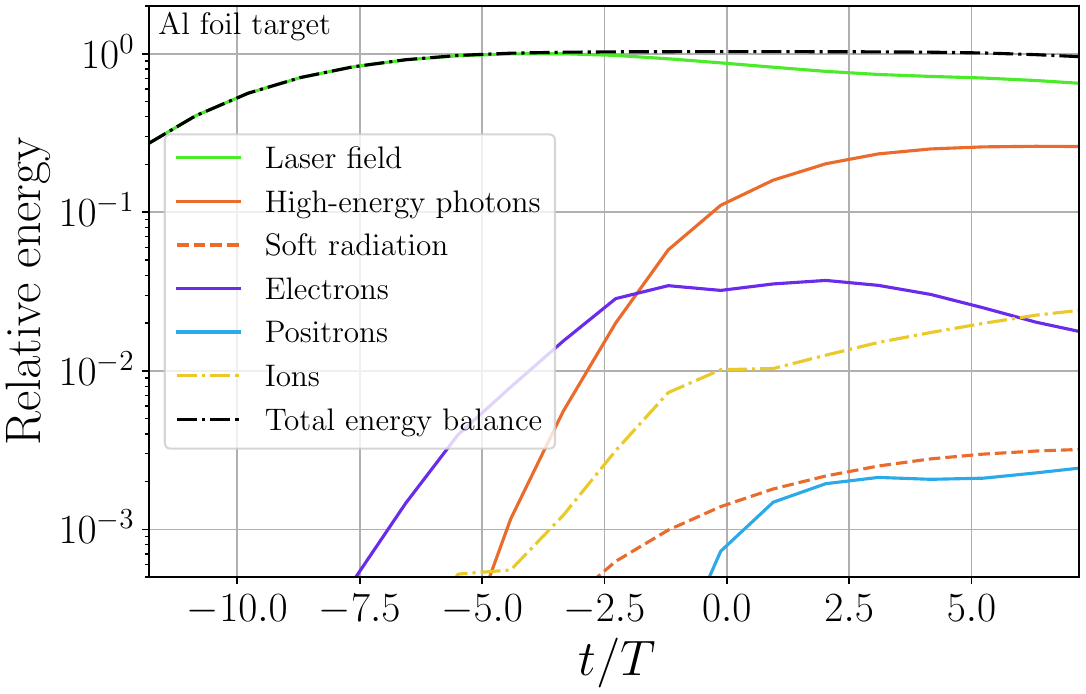}
	\caption{Time dependence of the energy balance between the laser field, radiation, and particles. The results are obtained from PIC simulations with an aluminium foil target (see also Tab.~\ref{tab:seeding}). The data is normalized to the total energy of two laser pulses ($\approx$1000 J). The orange curve is obtained by integrating the spectral data for photons with energy $>1$ MeV. The laser field configuration is described in Section~\ref{sec:iiib}. \CORR{Note that the figure shows the total energy contained in the actual simulation box. The maximum laser energy (as well as the energy balance) is reached at $t\gtrsim -5T$, when the pulses fully enter the simulation box.}}
	\label{fig:energy_balance}
\end{figure}

A significant number of $e^-e^+$ pairs are created during the interaction for all targets considered in our simulations. In Figs.~\ref{fig:seeding}(a-3), (b-3), and (c-3), we plot the time dependence of the positron number $N_{e^+}$, normalized by $\Nsf$ for each target. Pair creation approximately starts when the first seed electrons gain $\chi_e\sim 1$, which occurs around $t\approx-3T$ [see the corresponding curves for $\max(\chi_e^{\rm seed})$ in Figs.~\ref{fig:seeding}(a-4), (b-4), and (c-4)]. We associate this moment with the avalanche onset.

All three targets are suitable for triggering avalanches, and we find that the number of produced positrons provides a strong signature of the avalanche precursor. Notably, a steady exponential growth was still not reached for the chosen parameters. For this, higher fields or longer pulses are required. As we show in Section~\ref{sec:iv}, one of the options is using tighter focusing, e.g., $f/\sqrt{2}$.

The resulting total number of positrons is presented in Tab.~\ref{tab:seeding}. Our simulations predict the creation of approximately one positron per seed electron in the focus for the krypton and plasma targets. The number $N_{e^+}/\Nsf$ is somewhat lower for the foil target. This is due to partial screening of the laser field inside the foil due to its high density [see Fig.~\ref{fig:seeding}(c-1)]. On the other hand, among the three targets, the total positron yield is the highest with the foil, which is due to a high number of seed electrons.

Let us also compare the energy transfer from the laser field to the hard photon radiation and charged particles. The corresponding values estimated at the end of the interaction\footnote{Hereinafter, we fix the particle distributions at $t\approx11 T$ as ``final''.} are presented in Tab.~\ref{tab:seeding}. As one can see, the percentage of the transferred energy strongly depends on the initial target density. For the foil, the total energy transfer from the laser field reaches $\sim30$\%, and most of it is contained in photons. Fig.~\ref{fig:energy_balance} shows the evolution of the species energy during the interaction with the aluminium foil. As electrons are accelerated by the field, they rapidly start losing the gained energy in emission, which dominates the particle dynamics, namely, it takes place deeply in the RD regime. The radiative, i.e., the high-energy photons, component still dominates after the cascade onset and its further development.

\subsection{Final particle distributions}
\label{sec:iiie}
\begin{figure}
	\centering
	\includegraphics[width=\linewidth]{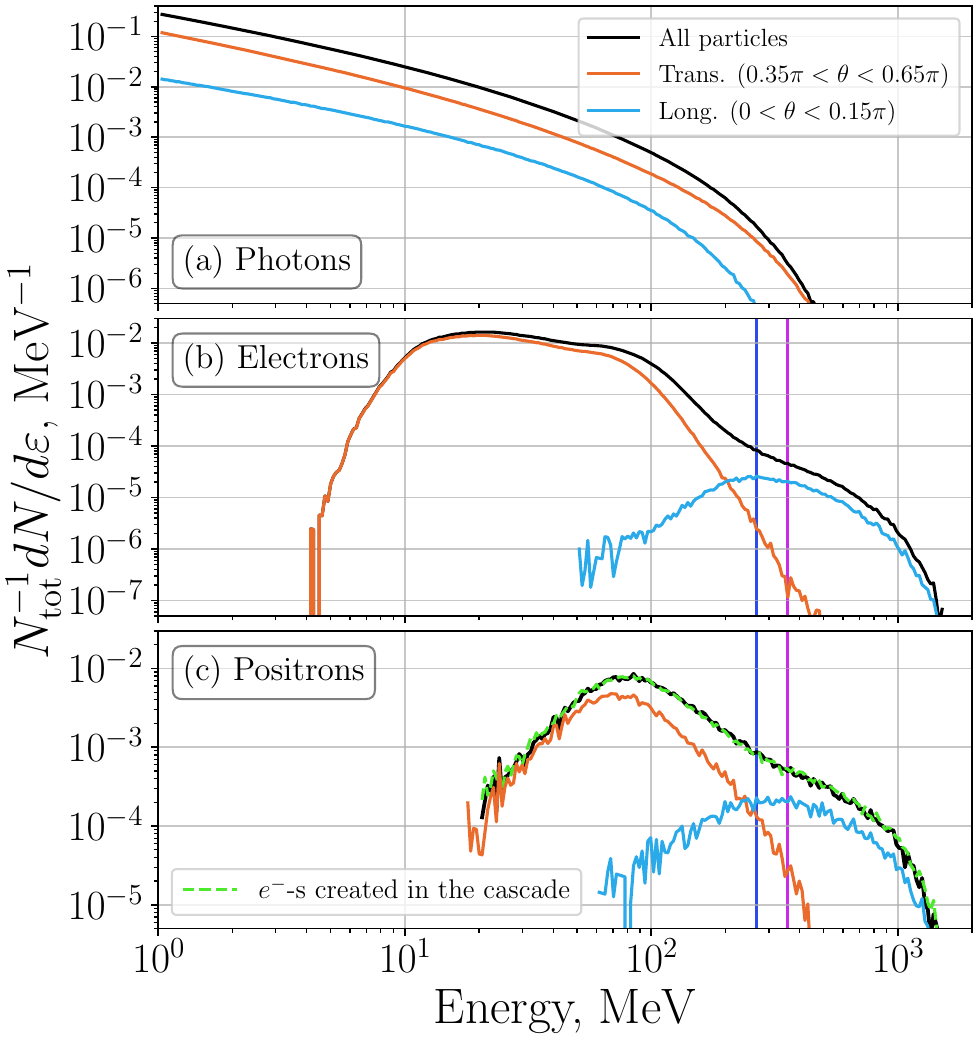}
	\caption{The final spectra of (a) photons, (b) electrons, and (c) positrons. In all plots, the black line shows the spectrum calculated for all particles in the simulation box, the orange line corresponds to particles propagating in the polar angle opening $0.35\pi<\theta<0.65\pi$, the blue line to particles propagating in $0<\theta<0.15\pi$ (recall that $\theta=0$ corresponds to the laser propagation axis). The green dashed line in (c) shows the spectrum exclusively for electrons that were created in the cascade. The vertical dark-blue and magenta lines show the maximum $e^-$ energy estimated from Eq.~\eqref{energy_cl_LL} and $\gem mc^2$ given by Eq.~\eqref{tem_gem_cem}, respectively (see also Fig.~\ref{fig:landscape}). All spectra are normalized to the total number of the corresponding species. The injection is implemented via a krypton gas (see Tab.~\ref{tab:seeding}), and the laser field configuration and parameters are presented in Section~\ref{sec:iiib}.}
	\label{fig:spectra}
\end{figure}
\begin{figure*}
	\centering
	\includegraphics[width=0.329\linewidth, trim={35 50 10 60}, clip]{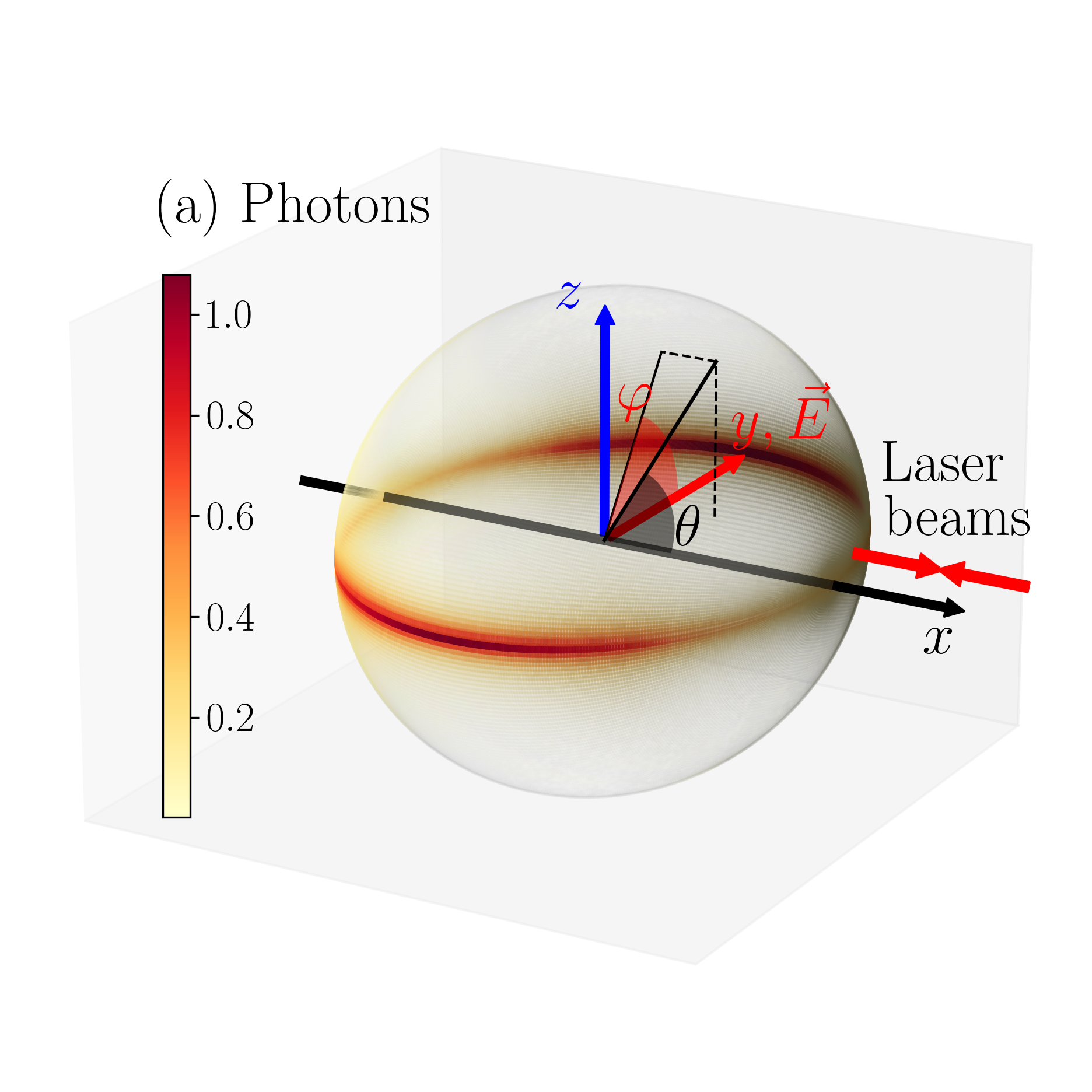}
	\includegraphics[width=0.329\linewidth]{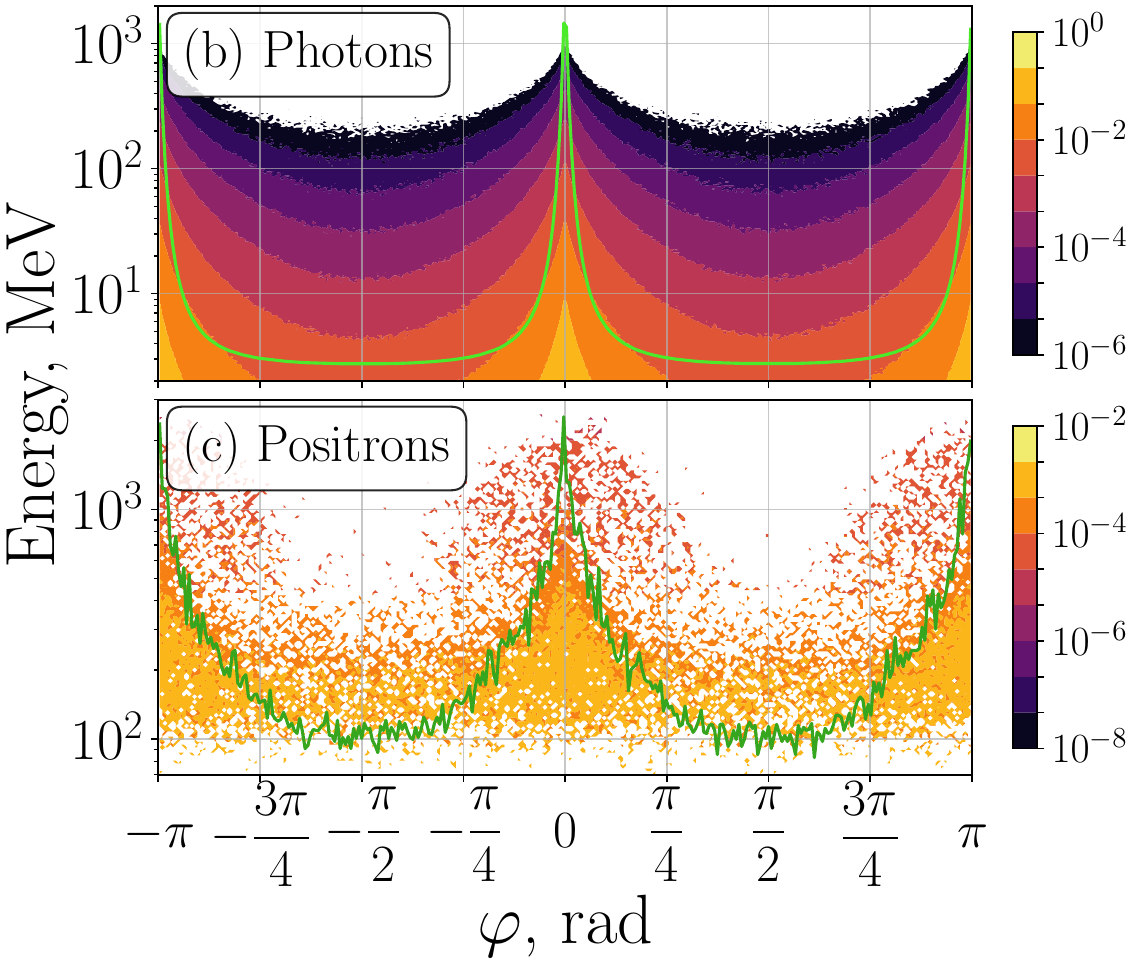}
	\includegraphics[width=0.329\linewidth]{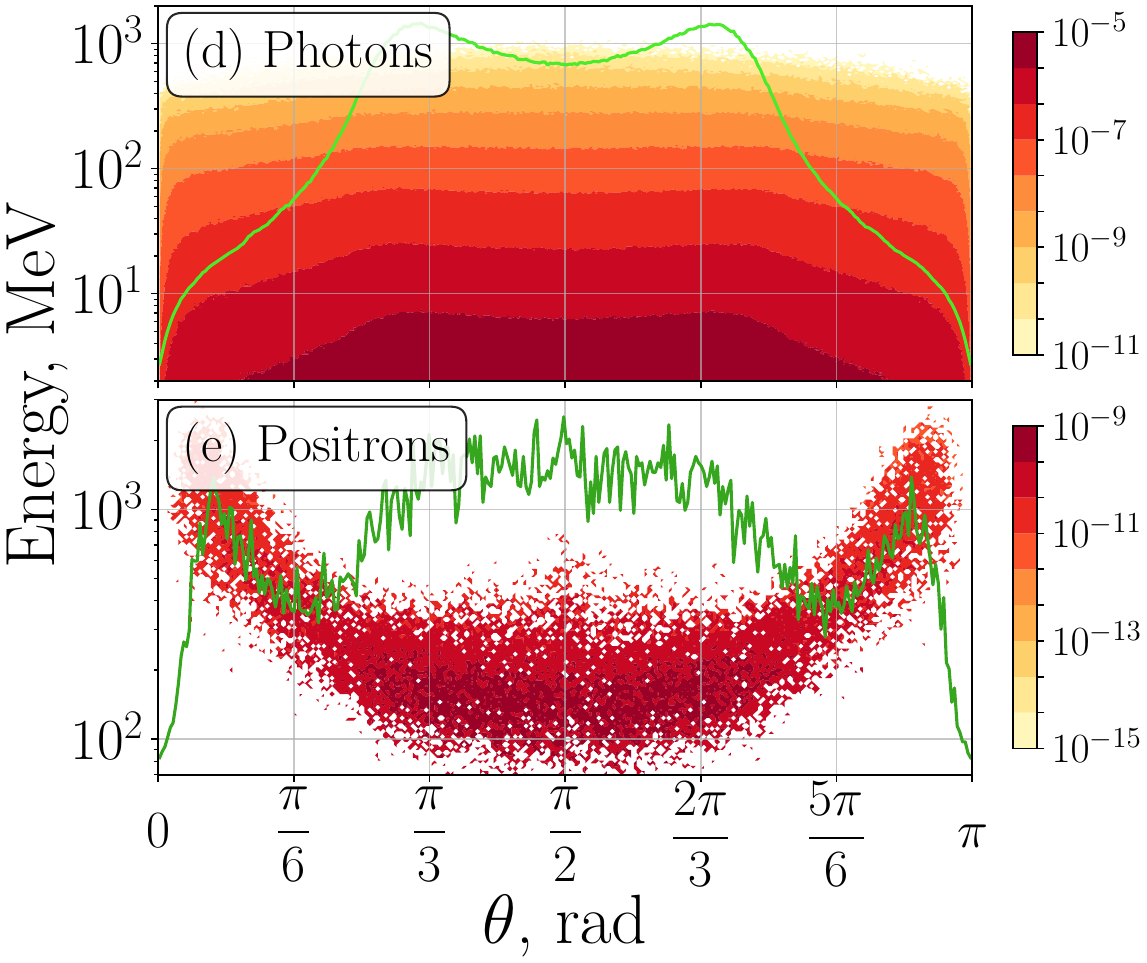}
	\caption{(a) Final angular distribution of photons (in arb. units). Note that the electric field is polarized linearly in the $y$-direction ($\theta=\pi/2$, $\varphi=0$ or $\pi$). (b-e) Final photon and positron angular-energy distributions (normalized to unity). The parametrisation of angles is illustrated in (a), which is used throughout. The green curves show the energy-integrated angular distribution on a linear scale in arb. units. (namely, the 2D plot integrated along the \CORR{vertical axis}). Here, the target is krypton gas (see Tab.~\ref{tab:seeding}), and the laser field configuration and parameters are presented in Section~\ref{sec:iiib}. }
	\label{fig:spectra_angular}
\end{figure*}

The final energy and angular distributions resulting from the interaction are fundamental quantities of interest for experiments. The normalized final particle distributions resulting from the three considered targets share similar features. Therefore, in this subsection, we provide the simulation data obtained for the krypton gas. The final particle spectra and angular-energy distributions are presented in Figs.~\ref{fig:spectra} and \ref{fig:spectra_angular}.

Let us first consider the full spectra of electrons and positrons [see the black lines in Figs.~\ref{fig:spectra}(b) and (c)]. The electron spectrum is dominated by lower energies, as compared to the positron distribution. This is due to the contribution from the initial electrons. As they are distributed in the volume, a significant fraction of these electrons is exposed to weaker fields, so they gain less energy during the interaction. The spectrum for the electrons that were created in the cascade [see the green dashed line in Fig.~\ref{fig:spectra}(c)] matches the positron spectrum perfectly.

The high-energy behaviour of the $\gamma$ and $e^\pm$ spectra is linked to the geometry of the interaction. Most $e^\pm$ in the focus of the standing wave are accelerated transverse to the laser propagation axis by the electric field of the standing wave. However, the pulses can accelerate some of the charged particles longitudinally, i.e., along the laser propagation after the crossing.\footnote{Electrons injected a (single) laser pulse with $a_0\gg1$ rapidly reach ultra-relativistic velocities and get ``phase locked'', allowing them to co-propagate with the laser wave,  i.e. the customary eight-figure for a moving electron in a LP wave is stretched.\cite{landau2013classical} This is analogous to the vacuum laser acceleration.\cite{wang2001vacuum} This results in acceleration to high energies in the longitudinal direction. Also note that positrons propagating longitudinally are less affected by the RR.} 

We now consider the angular dependence of the spectra from this perspective. Figure~\ref{fig:spectra_angular} shows the final angular-resolved distributions.
\CORR{The angles are chosen with the following convention (illustrated in Fig.~\ref{fig:spectra_angular}a): $\varphi$ is the azimuthal angle, $-\pi<\varphi\leq\pi$, counted from the positive direction of the $y$-axis in the $(yz)$-plane, and $\theta$ is the polar angle, $0\leq \theta<\pi$, counted from the positive direction of the $x$-axis.}  
Recall that the laser pulses propagate along the positive or negative $x$-direction ($\theta=0$ or $\theta=\pi$, respectively) and the electric field is polarized in the $y$-axis ($\theta= \pi/2$, $\varphi=0$ or $\pi$). For brevity, we refer to these directions as longitudinal and transverse, respectively.

As seen from Figs.~\ref{fig:spectra_angular}(a), (b), and (d), photons are predominantly emitted transversely in the $xy$-plane, and the most energetic photons are distributed around the transverse direction. For illustration, in Fig.~\ref{fig:spectra}(a), we also show the spectra for photons emitted around a selected direction with the angular spread $\Delta\theta=0.15\pi$: longitudinally, so that $0<\theta<\Delta\theta$ (the blue curve), and transversely, $\pi/2-\Delta\theta<\theta<\pi/2+\Delta\theta$ (the orange curve; \CORR{see also Fig.~\ref{fig:scheme} for illustration). Note that the value $\Delta\theta=0.15\pi$ is chosen for illustration, and variation of $\Delta\theta$ does not affect the overall discussion presented below as long as the transverse and longitudinal motion can be separated (see Appendix~\ref{app:theta} and Fig.~\ref{fig:spectra_delta_theta} therein for more details).}

The spectra of charged particles carry additional information about the interaction dynamics. As follows from the $\varphi$-distribution in Fig.~\ref{fig:spectra_angular}(c), charged particle motion happens mostly parallel to the $xy$-plane, which also explains the photon emission pattern. The $\theta$-distribution in Fig.~\ref{fig:spectra_angular}(e) shows that a fraction of the positrons are accelerated longitudinally in the fields of the pulses(see the lobes rising near $\theta\sim 0$ and $\theta\sim \pi$). The longitudinal acceleration explains the extension of the high-energy tail beyond 1 GeV in the electron and positron spectra shown in Figs.~\ref{fig:spectra}(b) and (c), respectively. Particles that travel transversely have a cut-off at noticeably smaller energies around $\sim300\text{-}400$ MeV. 

Let us consider the high-energy tails for the transversely propagating particles in more detail [see the orange curves in Figs.~\ref{fig:spectra}(b) and (c)]. 
We can compare the energy of the tails to the estimates based on the RR models introduced in Sections~\ref{sec:iia} and \ref{sec:iib}. The classical expression with the LL term, given by Eq.~\eqref{energy_cl_LL},\footnote{As we mentioned, Eq.~\eqref{energy_cl_LL} is derived for a rotating electric field, but it can be used in estimates for other field configurations.} predicts a maximum electron energy $\varepsilon_e=mc^2\gamma_e\approx 270$ MeV for $\atot=1100$ and $\lambda=0.91$ $\mu$m. We mark this threshold by a blue vertical line in Figs.~\ref{fig:spectra}(b) and (c). As one may see, the exponential tails extend beyond the prediction of the classical theory, hence, as expected, the RR is manifestly quantum. The exponential cut-off location can be estimated with Eq.~\eqref{tem_gem_cem}, which gives $\gem mc^2\approx 360$ MeV (also shown in the figure with the magenta line). This is coherent with our simulation results.

Summarizing, most of the radiation and new particles generated in the cascade are ejected in the direction of the electric field polarization axis. The spectra measured at the corresponding angles \CORR{can be used} to characterize the interaction in the focal center.

\section{Parametric study}
\label{sec:iv}
\begin{figure}
    \centering
    \includegraphics[width=\linewidth]{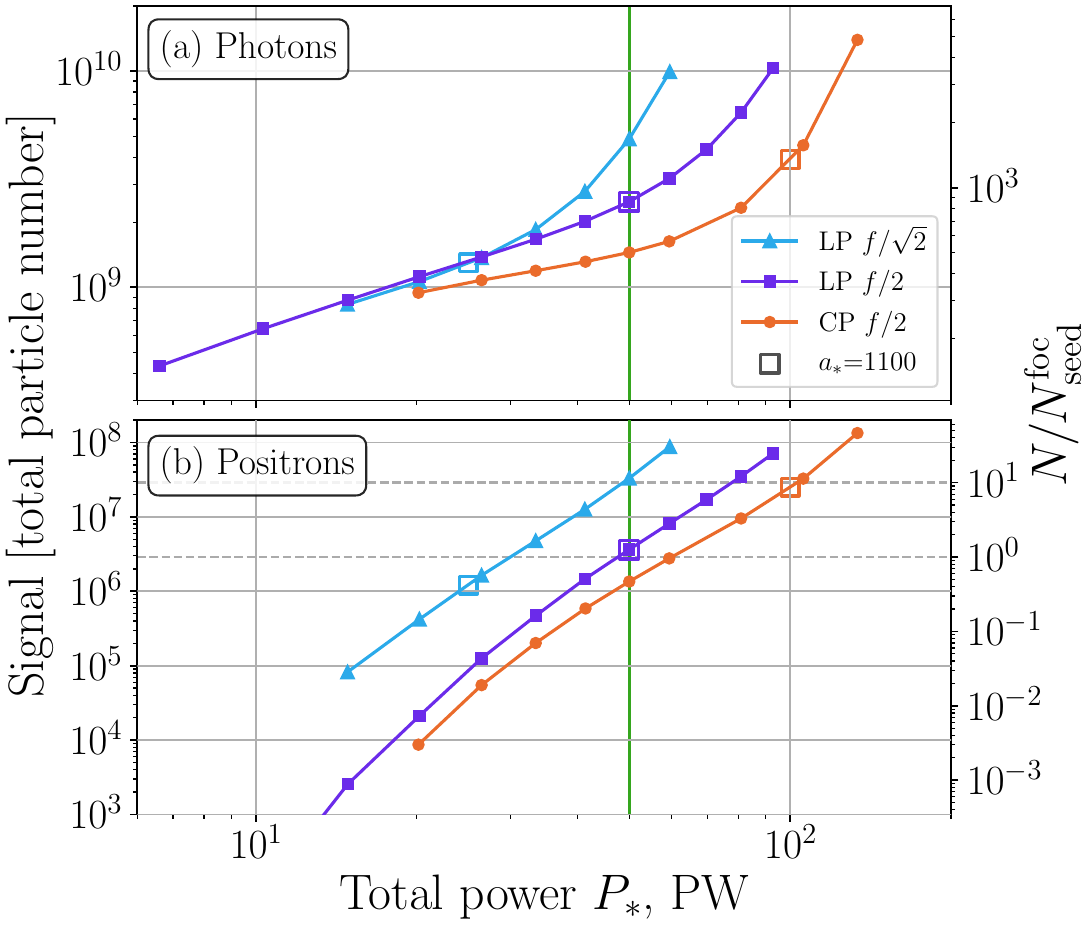}
    \caption{The total number of photons and positrons as functions of the total laser power $\Ptot$ for different focusing and polarization. An empty square in each curve corresponds to the signal calculated at peak field strength $\atot=1100$. The data is obtained from 3D PIC simulations with a reduced electronic target (see Section~\ref{sec:iva}). The vertical axis on the l.h.s. shows the total particle numbers assuming the initial electron density $n\approx 3.97\times 10^{18}$ cm$^{-3}$; the vertical axis on the r.h.s. corresponds to the relative particle numbers normalized to the number of seed electrons $\Nsf\approx2.9\times10^6$. The green vertical line corresponds to $\Ptot=50$ PW.}
    \label{fig:total_N}
\end{figure}

\begin{figure}
	\centering
	\includegraphics[width=\linewidth]{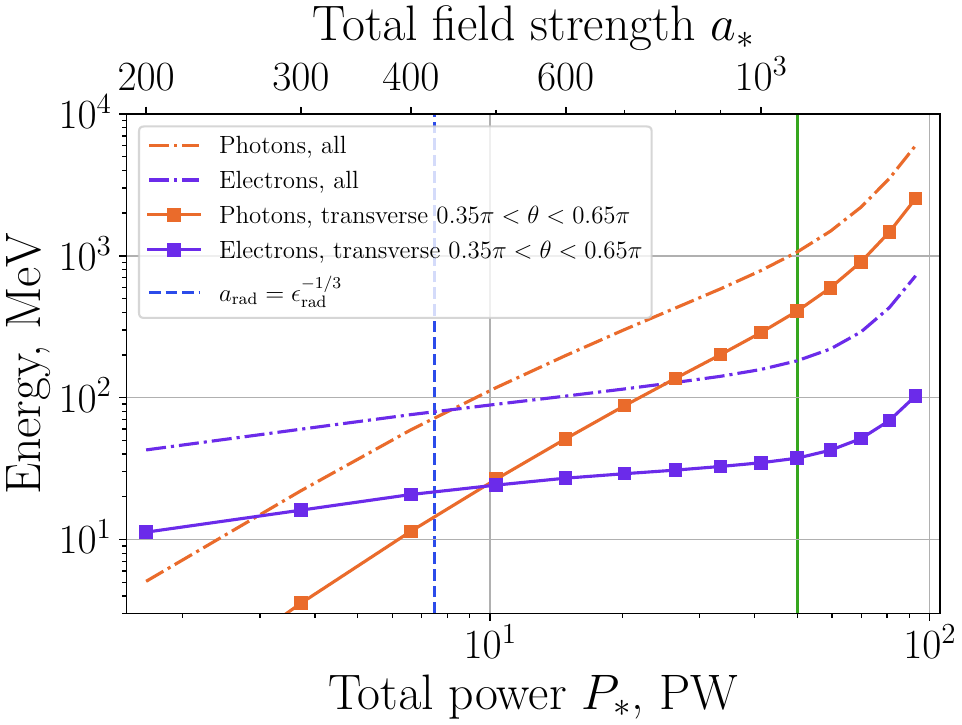}
	\caption{Final total energy of photons and electrons as a function of the peak total power (or the total field strength, shown on the upper horizontal axis) for the case of LP $f/2$. \CORR{The plotted energy is per seed electron ($\Ns\approx4\times 10^7$).}}
	\label{fig:energy} 
\end{figure}

So far, we have considered a fixed laser field setup. Let us now study how the resulting number of particles and their energy depend on the field parameters. As we have discussed in Section~\ref{sec:iib}, the variation of the peak field strength $\atot$ from lower to higher values allows one to study the transitions among different interaction regimes. In this section, we also study the generation of cascades with a CP field configuration and compare it to the LP case. Finally, we consider the effect of tighter focusing (in the LP case). 

The peak field strength $\atot$ is the key parameter, which controls the interaction. However, in practice, it is varied by changing the total power $\Ptot$. Therefore, in our studies, we choose $\Ptot$ as the main parameter. Note that at fixed $\Ptot$, the values of $\atot$ and intensity depend on the field polarization and focusing. Then, for the cases we consider, the peak field strengths are related as $\atot {}_{\mathrm{LP},f/\sqrt{2}}=\sqrt{2}\atot {}_{\mathrm{LP},f/2}$ and $\atot {}_{\mathrm{CP},f/2}=(1/\sqrt{2})\atot {}_{\mathrm{LP},f/2}$, where the indices denote the polarization and focusing degree.

\subsection{Idealized seed electron target}
\label{sec:iva}
For this study, we use an idealized target for injection of electrons in the focus. We assume that it consists only of electrons homogeneously distributed in a ball of radius $1.34$ $\mu$m. In our simulations, we build on the setup from Sec.~\ref{sec:iiib} and vary its parameters. To speed up the numerical computation, we start from the time $t=-0.5T$, assuming that the laser fields have already arrived at the interaction point (recall that the maximum electric field reaches the origin $\vec{r}=\vec{0}$ at $t=0$). The electrons are injected in the field at the start of the simulation. Finally, we set the particle density to $n=3.97\times10^{18}$ cm$^{-3}$. With this density, the number of seed electrons in the small focal volume $\lambda^3$ matches $\Nsf$ for the krypton target. \CORR{The total number of seed electrons contained in the idealized target is $\Ns\approx 4\times10^7$.}
 
We performed a test simulation at the field parameters that we used to obtain the results for krypton shown as black solid curves in Fig.~\ref{fig:seeding}(a). We plot the evolution of the seed electron number in the focus for the idealized electron target in Fig.~\ref{fig:seeding}(a-2) with the black solid curve. As seen from the figure, the curve follows the results for krypton. Also, we plot the total number of positrons produced in the cascade per seed electron in the focus in Fig.~\ref{fig:seeding}(a-3). The resulting pair yield for the realistic and idealized targets match, as well as the maximum values of $\chi_e$, shown in Fig.~\ref{fig:seeding}(a-4). Therefore, we conclude that the idealized target can be used for the parametric studies.

\subsection{Total particle number and energy dependence on the laser parameters}
\label{sec:ivb}
Let us consider the dependence of the total number of positrons and high-energy ($>1$ MeV) photons (for brevity, the signal) produced during the interaction. In Fig.~\ref{fig:total_N} we show its dependence on the total power $\Ptot$ delivered by two laser pulses. Note that while we reference the electron density to the krypton gas target, these results can be rescaled (at least within the range of densities for which collective effects are negligible), therefore, it is convenient to normalize the signal by $\Nsf$.

As one can see from Fig.~\ref{fig:total_N}, avalanche precursors with the yield at the level of $0.1\text{-}1$ $e^-e^+$ pair per seed electron in the focus are accessible in the range $P_*=30\text{-}50$ PW with $f/2$ focusing.\footnote{We conclude this assuming a perfect-case scenario for the laser beam setup alignment. However, as we discuss below in Section~\ref{sec:v}, avalanche precursors are also accessible in this parameter range in realistic geometries too.} 
For higher $\Ptot$, our simulation results show a rapid increase of the positron and photon numbers.  The transition to a prolific cascade is marked by the exponential dependence of the particle yield on $\Ptot$, which is seen in the photon signal. 

Tighter focusing allows for higher peak field strengths at the cost of a smaller interaction volume. We tested the case of $f/\sqrt{2}$ focusing (LP) and found that a higher particle yield is obtained in the full range of $\Ptot$ that we considered. It exceeds the results for $f/2$ (LP) by an order of magnitude. Furthermore, at $f/\sqrt{2}$, the exponential increase of the signal with the field strength is already seen at $\Ptot>20$ PW. Thus, with the capacity of $P_*= 50$ PW, it is potentially feasible to reach not only an avalanche precursor, but also a prolific cascade and access the green area in Fig.~\ref{fig:landscape}. 

As for the polarization, our simulations show that the particle yield is lower for the CP case at fixed $\Ptot$ and focusing. However, let us emphasise that at fixed \textit{field strength} $\atot$, the conclusion is the opposite: the CP wave generates a higher signal at the same focusing. We illustrate this in Fig.~\ref{fig:total_N} for the case of $\atot=1100$ by the points marked with empty squares. In the CP field configuration, the electric field amplitude does not oscillate as in the LP wave, meaning that the particles are exposed to a strong electric field during the whole duration of the laser pulses. This allows the interaction to reach the exponential phase of the cascade faster, which results in a higher multiplicity.

The transitions between the interaction regimes can be clearly distinguished in the field strength dependence of the total energy carried by photons and electrons after the interaction. We plot this quantity in Fig.~\ref{fig:energy} for the case of LP and $f/2$ focusing for (i) all particles in the simulation box and (ii) particles that propagate along the electric field polarization axis in the angle opening $\pi/2-\Delta\theta<\theta<\pi/2+\Delta\theta$, $\Delta\theta=0.15\pi$ (as with the orange curves in the spectral plots in Fig.~\ref{fig:spectra}).\footnote{\CORR{We use the following procedure to calculate the total energy carried by particles moving in a given interval of angles $\theta_1<\theta<\theta_2$. We extract the energy-angular distribution $f(\varepsilon,\theta)$ from simulations at a time point after the interaction. Given this distribution is normalized such that the integration gives the number of particles contained in the chosen part of the phase space, $N = \int_{\theta_1}^{\theta_2} d\theta\, \sin\theta\int d\varepsilon\, f(\varepsilon,\theta)$, the energy carried by these particles can be calculated as $\varepsilon_{\mathrm{tot}}= \int_{\theta_1}^{\theta_2} d\theta\, \sin\theta\int  d\varepsilon\, \varepsilon f(\varepsilon,\theta)$. In Fig.~\ref{fig:energy}, we plot $\varepsilon_{\mathrm{tot}}/\Ns$, namely, energy per seed electron.}}

Both the energy of photons and electrons grow rapidly with $\atot$. The first notable feature is the dominance of the photon energy at $\atot>a_{\mathrm{rad}}$, as the accelerated electrons emit a lot of radiation. This is a strong signature of the transition to the RD regime, and it is consistent with the estimates in the classical picture [see Eq.~\eqref{energy_cl_LL}]. The transition is clearly seen from $\Ptot\approx8 - 10$ PW both when energy is calculated for all particles, and for particles moving transversely to the laser propagation axis. As the field is increased further, the energies grow monotonically. The transition to the avalanche precursor is better seen in the positrons signal, which becomes noticeable at $\gtrsim 30$ PW [see the purple curves in Fig.~\ref{fig:total_N}]. At $\Ptot\gtrsim 30$ PW, the energies start to grow exponentially with the field strength, as the cascade becomes prolific and the particle numbers grow accordingly. 

To summarize, an analysis of the particle numbers, energies and spectra as a function of the laser power and intensity can be used to identify the regime of interaction. Furthermore, a study of the distributions for particles moving in the direction of the electric field polarization \CORR{can be used to analyse the interaction at the focal center. For example, the spectral cut-off location in this distribution can be employed to discriminate between the classical and quantum RD regimes. The dependence of the signal (from particles collected in the transverse direction) on the field intensity can be used to identify the transition to an avalanche. Finally, the shape of the spectra can also be used to deduce the cascade parameters, such as energy and $\chi$ of particles, by fitting analytical models.\cite{nerush2011analytical}}

\section{Realistic laser pulse alignment and desynchronization}
\label{sec:v}
\begin{figure}
	\centering
	\includegraphics[width=\linewidth]{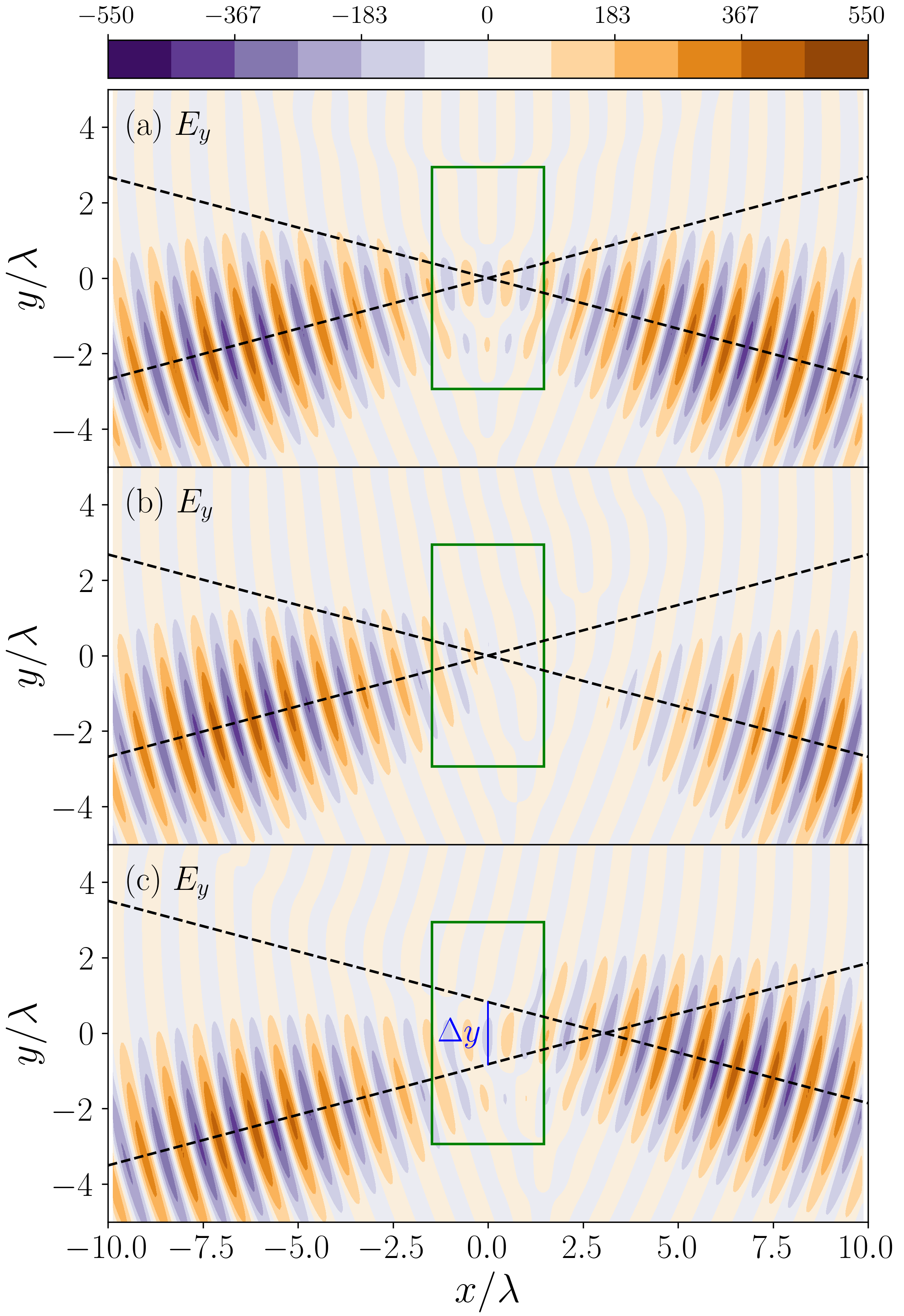}
	\caption{\CORR{Snapshot of the $E_y$ field component distribution in a plane $z=0$ before the interaction for laser pulses that are: (a) tilted at 15$^\circ$ each, (b) tilted and one of the laser pulses is delayed, (c) tilted and focused with an offset $\pm\Delta y/2$ in the $y$ direction ($\Delta y\approx1.65\lambda=1.5$ $\mu$m). The dashed lines correspond to the propagation axes of the laser pulses. The green rectangle shows the initial shape of the krypton target used in simulations.}}
	\label{fig:field2}
\end{figure}

\begin{table}[]
	\caption{The dependence of the positron yield and $\chi_e$ of seed electrons on the non-ideality of the laser setup: beam tilting, mismatch of the beams' focal centers (misfocus) and a delay between the laser pulse arrival at the interaction point (mistiming). The target used in simulations is a flat slab of krypton of width $d=0.75w_0\approx 1.5\lambda$. For tilting, the laser beams are rotated around the $z$-axis (parallel to the $\vec{B}$ field component). The angle of $30^\circ$ between the laser pulse propagation axes is chosen to ensure that the laser beams do not back-propagate through the generating optical systems after the interaction (recall that the lasers are assumed here to be focused to $f/2$). The parameters for misfocusing and mistiming are chosen in accordance with Table~\ref{tab:nsf_opal_params}. Note that $\langle\chi_{\rm seed}\rangle$ is a function of time [e.g., see the dark-blue line in Fig.~\ref{fig:seeding}(a-4)], and in the table we present its maximum value.}
	\label{tab:misalignment}
	\centering
	\begin{tabular}{ccc| c c c}
		\hline\hline
		Tilting & Misfocus & Mistiming & $N_{e^+}$ & $\max \langle \chi_{\rm seed}\rangle$ & $\max\chi_{\rm seed}$   \\
		\hline
		- & - & - & 1 & 0.75 & 4.2\\
		$30^\circ$ & - & - & 0.81 & 0.64 & 3.6 \\
		- & $\Delta y=1.5$ $\mu$m & - & 0.48 & 0.65 & 3.3\\
		- & $\Delta z=1.5$ $\mu$m & - & 0.37 & 0.64 & 2.8\\
		- & - & 10 fs & 1.61 & 0.75 & 3.1\\
		$30^\circ$ & $\Delta y=1.5$ $\mu$m & - & 0.48 & 0.57 & 2.8\\
		$30^\circ$ & $\Delta z=1.5$ $\mu$m & - & 0.27 & 0.53 & 2.8\\
		$30^\circ$ & $\Delta z=1.5$ $\mu$m & 10 fs & 0.21 & 0.45 & 2.0\\

		\hline\hline
	\end{tabular}
\end{table}

Finally, let us demonstrate the stability of the signals to the effect of a more realistic laser setup layout. For example, the axes of the counterpropagating beams should be tilted to avoid possible back-propagation of intense light in the optical systems after the interaction. Non-ideal co-pointing and co-timing can reduce the maximum field reached at the focus, therefore reducing the signal. We study the impact of these featuress within the parameter range suggested in Tab.~\ref{tab:nsf_opal_params}.

Recall that in the ideal case, the laser pulses exactly counterpropagate with respect to each other along the $x$-axis, are polarized linearly in the $y$-direction and focused to $f/2$ (as in Section~\ref{sec:iii}). Then, the laser propagation axes are tilted in the $(xy)$-plane, so that the relative angle between them amounts to 30$^\circ$ (in simulations, we tilt each beam by 15$^\circ$, as this angle allows avoiding back-propagation at $f/2$ focusing). To account for possible mismatch of the focal centers (for brevity, ``misfocus''), we shift their position in the $(yz)$-plane at $x=0$. We tested two cases, when the focal centers are shifted in the $y$- or $z$-direction by 1.5 $\mu$m. Finally, we assume that one of the pulses can be delayed by 10 fs (``mistiming''). \CORR{Fig.~\ref{fig:field2} illustrates different field configurations with tilted, delayed, and misfocused laser pulses.}

In this set of simulations, to treat the electron injection realistically, we consider a Kr$^{26+}$ gas target. It is shaped as a rectangular block sized \CORR{$d_x=1.5w_0=2.68$ $\mu$m along the $x$-axis, and $d_y=d_z=2d_x$} in the transverse directions,\footnote{This change from the ball-shaped considered in Sections~\ref{sec:iii} and \ref{sec:iv} is motivated by the convenience for studying the effects of misfocusing.}\CORR{which is also shown in Fig.~\ref{fig:field2}.} As in Section~\ref{sec:iii}, at the start of each simulation, the laser pulses are injected from the simulation box sides and propagate to the center, ionize the target and trigger a cascade. The simulation parameters are also the same as described in Section~\ref{sec:iiib}.

The results are summarized in Tab.~\ref{tab:misalignment}. We calculate the total number of positrons produced and $\chi_e$ reached by seed electrons during the interaction. For convenience, the results for $N_{e^+}$ are gauged to the ideal case. First, we test the three modifications of the setup one by one. The expected suppression of the positron signal is caused by the reduction of the field in the focus (e.g. seed electrons get lower average and maximum values of $\chi_e$). The exceptional case is mistiming, for which we find an even higher signal. In this case, the first pulse not only ionizes the target before the maximum field is reached, but also pre-accelerates the seed electrons. Hence, a faster onset of the cascade is possible.\cite{mironov2014collapse} This suggests some possibility for further optimization of the setup. 

The laser beam axes tilting is the most restrictive modification of the setup, as it is imposed by the facility design requirements. Fortunately, we find that its effect on the pair yield is minor (at least for the parameter range that we considered). The signal is more sensitive to misfocusing. However, assuming that it can be caused by the laser beam shot-to-shot jitter, this issue is mitigated by repeating the shots and improving the statistics. The same applies to mistiming. Notably, a gas target is suitable for such repetitive experiments.

\section{Conclusions}
\label{sec:vi}
We studied the possibility of generating avalanche-type QED cascades with two counterpropagating laser pulses in the range of the NSF OPAL parameters (summarized in Tab.~\ref{tab:nsf_opal_params}) and showed that this future facility is uniquely placed for such experiments. The two-beam extremely high power laser configuration offers the possibility to study SFQED phenomena that can not be accessed by any other existing or planned laser facility.  We employed 3D SMILEI PIC simulations, which allow for a consistent treatment of strongly-focused laser pulses and particle dynamics, including collective effects and the relevant quantum processes of nonlinear Compton emission, Breit-Wheeler pair production, and tunneling ionization. 

We considered several realistic configurations for the injection of electrons to trigger cascades. We went beyond an ideal laser field configuration by considering an interaction geometry with tilted laser beams, which prevents back-propagation of light in the optical systems, and studied the effect of non-ideal focusing. Our simulations show that QED cascades can be reliably generated under realistic conditions, which is revealed in strong positron and high-energy photon signals. 

The most stringent requirements for the two-pulse configuration are the alignment of the electric field polarization axes, envelope co-timing, and co-location of the focus on the target. While there are potentially other non-ideal effects, such as the pulses having different carrier-envelope phases or target inhomogeneities, we expect these to have less impact on the cascade than the effects investigated here.

The variation of the parameters below and beyond those planned for NSF OPAL shows that almost all of the matter-radiation interaction regimes charted in Fig.~\ref{fig:landscape} are within reach. This includes the transition from classical to quantum radiation, the generation of bright gamma flashes in the RD regime, and, at peak powers $\Ptot\approx40\text{-}50$ PW, the production of avalanche precursors. With a stronger focusing ($f/\sqrt{2}$), it is feasible to generate prolific cascades with an order-of-magnitude higher multiplicity. Observation of all these regimes and transitions between them would be a landmark test of strong-field QED. Prolific avalanches would also open a path to the creation of a relativistic electron-positron plasma in a laboratory.\cite{mercuri2025growth}

Our simulation results can be used to guide the design of future experiments. Electrons can be reliably injected into a strong field using a static target. This can be a heavy-atomic gas such as krypton, a plasma slab, or a thin foil. Each target has advantages. For example, the total positron and photon yield depends on the initial target density. Therefore, solid thin foils are useful for generating stronger signals. However, as the foil thickness is comparable to the laser wavelength, it requires a high-precision placement in the focal region. This may impose additional restrictions on the setup design.

Heavy-atomic gas does not suffer from this issue, as it can be distributed in a large volume. In this case, a cascade can be triggered at any location. Also, a gas target is suitable for high-repetition experiments, and therefore for collecting large statistics. Another advantage of the gas target is the possibility to use ions for an \textit{in situ} measurement of the laser field intensity.\cite{ciappina2020focal, ouatu_pre2022} Combined diagnostics for ions and accelerated electrons\cite{longman2023toward} can potentially help reconstruct the intensity profile at focus in each shot, which is beneficial for experiment interpretation.

The final particle numbers and energy distributions resulting from the interaction have a strong directional pattern. Particles and radiation coming from the laser focus propagate predominantly transversely to the laser optical axis. For an LP field, they are spread close to the electric field polarization axis, forming relativistic jets. This feature can be used in the detector layout design. 

We focused on proof-of-principle simulations, demonstrating the possibility of producing avalanche-type cascades reliably in realistic conditions. The considered scheme can be optimised by tuning the properties of the target and laser parameters. While the generation of avalanche-type cascades is of fundamental importance by itself, bright gamma flashes and relativistic electron-positron-photon jets produced in the process may also have relevance to astrophysical phenomena. Establishing this link in the context of the experimental potential of NSF OPAL would be of interest for future work. 

\section*{Acknowledgements}
We express our gratitude to J. D. Zuegel for discussions and support of this project. We are also grateful to Jake Bromage for providing useful information on the envisaged contrast at NSF OPAL.

This material is based upon work supported by the U.S. National Science Foundation Mid-scale Research Infrastructure Program under Award No. PHY-2329970. Any opinions, findings and conclusions or recommendations expressed in this material are those of the author(s) and do not necessarily reflect the views of the National Science Foundation.

All of the simulation results we obtained on the HPC system ``3Lab Computing'' hosted at \'{E}cole polytechnique and administrated by the Laboratoire Leprince-Ringuet, Laboratoire des Solides Irradi\'es and Laboratoire pour l'Utilisation des Lasers Intenses. S.S.B. was supported by U.S. Department of Energy Office of Science Office of High Energy Physics under Contract No. DE-AC02-05CH11231. A.D.P. is partially supported by the U.S. National Science Foundation Mid-scale Research Infrastructure Program under Award No. PHY-2329970. The work of G.G. was partially supported by UKRI under grants ST/W000903/1 and EP/Y035038/1. S. Meuren is supported by the Centre National de la Recherche Scientifique (CNRS) and the Agence Nationale de la Recherche (ANR) under the Chaire de Professeur Junior (CPJ) program. Financial support by the ANR (g4QED project, Grant No. ANR-23-CE30-0011) is acknowledged.

This material is based upon work supported by the U.S. Department of Energy [National Nuclear Security Administration] University of Rochester ``National Inertial Confinement Fusion Program'' under Award Number DE-NA0004144 and U.S. Department of Energy, Office of Science, under Award Number DE-SC0021057.

This report was prepared as an account of work sponsored by an agency of the United States Government. Neither the United States Government nor any agency thereof, nor any of their employees, makes any warranty, express or implied, or assumes any legal liability or responsibility for the accuracy, completeness, or usefulness of any information, apparatus, product, or process disclosed, or represents that its use would not infringe privately owned rights. Reference herein to any specific commercial product, process, or service by trade name, trademark, manufacturer, or otherwise does not necessarily constitute or imply its endorsement, recommendation, or favoring by the United States Government or any agency thereof. The views and opinions of authors expressed herein do not necessarily state or reflect those of the United States Government or any agency thereof.

\section*{Data availability}
The data that support the findings of this study are available from the corresponding author upon reasonable request.

\appendix
\section{The probability rates of the nonlinear Compton emission and Breit-Wheeler pair production}
\label{app:rates}
Within the locally constant field approximation,\cite{ritus1985,fedotov2023high} which we assume to hold with a noticeable margin for the parameters considered in this work,\cite{gelfer2022nonlinear}, the differential probability rate for an electron with the Lorentz factor $\gamma_e$ and quantum parameter $\chi_e$ to emit a photon with energy within the interval $mc^2\times(\gamma_\gamma, \gamma_\gamma+d\gamma_\gamma)$ is given by 
\begin{widetext}
	\begin{gather}
		\label{eq:rate_CS}
		\frac{d \WCS (\gamma_{e}, \chi_{e}, \gamma_\gamma)}{d \xi} = \frac{1}{\sqrt{3}\pi} \dfrac{\alpha}{ \tau_C \gamma_{e}}   \left[ \left(1-\xi+\frac{1}{1-\xi}\right) K_{2/3}(\mu)-\int_{\mu}^\infty d s K_{1/3}(s)   \right] ,
	\end{gather}
	where $\tau_C=\hbar/(mc^2)$ is the Compton time, and we defined $\xi=\gamma_\gamma/\gamma_e\equiv \chi_\gamma/\chi_e$, $\mu = 2\xi/[3\chi_{e}(1 - \xi)]$. $K_{n}$ are the $n$-th order modified Bessel functions of the second kind. The $\chi$-parameter is conserved in the process, $\chi_e=\chi_\gamma+\chi_e'$, where $\chi_e'$ corresponds to the scattered electron. 

The differential probability rate for a photon with $\gamma_\gamma$ and $\chi_\gamma$ to produce a pair of $e^-e^+$, so that either $e^-$ or $e^+$ gets energy within the interval $mc^2\times(\gamma_e,\gamma_e+d \gamma_e)$, reads:
	\begin{equation}
		\label{eq:rate_BW}
		\frac{d \WBW(\gamma_{\gamma}, \chi_{\gamma}, \gamma_e)}{d\zeta}  = \frac{1}{\sqrt{3}\pi} \dfrac{\alpha}{ \tau_C \gamma_{\gamma}}  \left[\left( \frac{\zeta}{1-\zeta}+\frac{1-\zeta}{\zeta}\right)K_{2/3}(\mu') -\int_{\mu'}^\infty d s K_{1/3}(s) \right], 
	\end{equation}
	where $\zeta=\gamma_e/\gamma_\gamma\equiv\chi_e/\chi_\gamma$, $\mu' = 2 \zeta/[3\chi_{\gamma} (1 - \zeta)]$, and $\chi_\gamma=\chi_e+\chi_e'$. Here, $\chi_e$ and $\chi_e'$ correspond to the electron and positron. 
	
\end{widetext}

The total probability rates $\WCS (\gamma_{e}, \chi_{e})$, $\WBW(\chi_{\gamma}, \gamma_{\gamma})$  are obtained by integrating out the final particle energies or, equivalently, $\xi$ and $\zeta$: 
\begin{gather}
	\label{eq:rate_CS_tot}
	\WCS(\gamma_{e}, \chi_{e})=\int_0^1 \frac{d \WCS (\gamma_{e}, \chi_{e}, \gamma_\gamma)}{d \xi}d \xi,\\
	\label{eq:rate_BW_tot}
	\WBW(\gamma_{\gamma}, \chi_{\gamma})=\int_0^1 \frac{d \WBW(\gamma_{\gamma},\chi_{\gamma},  \gamma_e)}{d\zeta} d\zeta
\end{gather}
The expressions of the total probability rates simplify in the low- and high-field asymptotic regions:
\begin{gather}
	\renewcommand{\arraystretch}{1.5}
	\label{eq:rate_CS_asympt}
	\WCS(\gamma_{e}, \chi_{e}) \approx \frac{\alpha }{\tau_C\gamma_e}\times\left\lbrace
	\begin{array}{ll}
		1.44\,\chi_e, & \chi_e\ll 1, \\
		1.46\,\chi_e^{2/3}, & \chi_e\gg 1,
	\end{array}\right.\\
	\renewcommand{\arraystretch}{1.5}
	\label{eq:rate_BW_asympt}
	\WBW(\gamma_{\gamma},\chi_{\gamma}) \approx \frac{\alpha}{\tau_C\gamma_\gamma} \times \left\lbrace
	\begin{array}{ll}
		0.23\,\chi_\gamma e^{-\frac{8}{3\chi_\gamma}}, & \chi_\gamma\ll 1, \\
		0.38\,\chi_\gamma^{2/3}, & \chi_\gamma\gg 1.
	\end{array}\right.
\end{gather}

The probability rates and the product spectrum width grow with the $\chi$-parameter of the incoming particle (if the Lorentz factor is fixed). On the other hand, the $\chi$-parameter is shared among the products in each quantum event, so its value decreases for subsequent particle generation. This distinguishes the avalanche-type cascades from the ``usual'' shower-like events. While showers triggered by high-energy particles eventually saturate when the input energy of the incident bunch is exhausted,\cite{bulanov2013electromagnetic,mironov2014collapse} in avalanches, the field restores the particle energy by continuous acceleration. This allows to keep the probability rates of the quantum events high and sustain the cascade.

\section{\CORR{Approximations in use for modelling elementary SFQED processes and avalanche-type cascades}}
\label{app:approximations}
\CORR{
The kinetic approach underlying PIC-MC simulations is based on the interplay of several approximations.\cite{gonoskov_pre2015} As the laboratory-frame fields that we consider are below the Sauter-Schwinger limit, $E\ll E_S$, we can describe particle dynamics semi-classically. At the same time, the studied fields are strong in the sense that $a_0\gg 1$ (for example,  $a_0\gtrsim 5\times 10^2$ per beam at NSF OPAL). We consider a standing wave, and the electric field in the magnetic node acts twofold: (i) it accelerates charged particles and provides high $\gamma_e$ at a time $<1/\omega$, and (ii) it induces the quantum processes. The most relevant ones are the photon emission and pair creation, and their formation scale is typically of the order $\sim O(\lambda/a_0)\ll\lambda$. This enables us to use the locally constant field approximation (LCFA), see the reviews \cite{Di_Piazza_2012,Gonoskov_2022,fedotov2023high,popruzhenko2023dynamics} for a detailed discussion, and also Refs.\cite{ritus1985,baier1968} Indeed, in the kinetic picture, a particle propagates along a classical trajectory until it undergoes a quantum process randomly.
}

\CORR{
Let us discuss the set of used approximations in more detail. First, recall that a QED cascade is a high-multiplicity process. As the first major simplification, it is customary to consider it as a chain of first-order processes, namely, the nonlinear Compton emission and Breit-Wheeler pair creation. Such factorization is well-justified for processes in high fields with $a_0\gg1$ because the formation length, which scales as $l_f\sim \lambda/a_0$ is much smaller than a laser period and then than the pulse length $c\tau$. Thus, a factorized process made of $n$ elementary processes scales as $(c\tau/l_f)^n\sim (c\tau a_0/\lambda)^n$, whereas a nonfactorized process is linear in $c\tau$ and thus roughly scales as $c\tau a_0/\lambda$.\cite{baier1968} Hence, at high $a_0$, the former channel dominates, which motivates the chain approximation.
}

\CORR{
Second, let us briefly discuss the LCFA. This is a common approximation for the emission and pair creation rates when modelling cascades. Within it, the rates are defined by the local values of the background field at the position of the incoming particle. Due to the relativistic effects, the field in the particle rest frame is approximated by a constant crossed field (CCF), $E=B$, $\mathbf{E}\perp\mathbf{B}$, which enters the probabilities $\WCSBW$ only through the dependence on $\chi_{e,\gamma
}$, see Eqs.~\eqref{eq:rate_CS}, \eqref{eq:rate_BW}. 
}

\CORR{
A rigorous formulation of the LCFA applicability conditions is a challenging problem in SFQED, and we are not aware of a derivation valid in a general background field. For photon emission and pair creation, the LCFA rates typically arise in the limit of high $a_0$ and $\gamma_e$, which was shown for a magnetic field,\cite{baier1968} a plane wave,\cite
{nikishov_jetp1964, DiPiazza2018_LCFA,di2019improved,ilderton2019extended,heinzl2020locally, king2020uniform} a short laser pulse,\cite{seipt2011nonlinear, ilderton2019exact, titov2020multi, kampfer_pra2021} focused beam,\cite{di2021wkb, nielsen2022high,adamo2025scattering} crystal field, \cite{Wistisen_2018_b,Wistisen_2019_b} or in a general time-dependent electric field.\cite{gelfer2022nonlinear, mironov2023locally}
As we consider the interaction at the magnetic node of a standing wave, the latter field model is relevant for our case. Following Ref.\cite{gelfer2022nonlinear}, for photon emission, the LCFA applicability requires:
\begin{enumerate}
	\item Locality --- $a_0\gg 1$ and $a_0^3\gg \chi_e \chi_e'/\chi_\gamma$. 
	\item The field is close to crossed in the incoming and outgoing particles' frame --- $\gamma^3\gg \chi_e \chi_e'/\chi_\gamma$.
\end{enumerate}
Notably, the LCFA is a general approximation for SFQED effects not specific to photon emission and pair creation, and could be used, e.g. to treat Sauter-Schwinger pair creation in standing waves,\cite{aleksandrov2019locally,kohlfurst2022sauter} however, the applicability conditions can be different.}

\CORR{Generally speaking, the corrections to the LCFA can be relevant, for example, at $a_0\approx 1-10$, and $\gamma_{e,\gamma}\sim 10^2$. Such corrections can be calculated numerically, e.g. for a plane wave background or a single laser pulse.\cite{di2019improved, blackburn2021local,blackburn2022higher,montefiori2023sfqedtoolkit} At higher $a_0$ and $\gamma_{e,\gamma}$, the LCFA can be considered universally robust for various background field models. 
}

\CORR{
When modelling the onset stage of avalanche-type cascades, it is customary to consider the seed electrons initially at rest injected in the magnetic node of a standing wave.  One could argue that for such initial conditions, the LCFA rates may require corrections when describing photon emission (see e.g. Refs.~\cite{raicher2015novel, king2016classical} for details). At $a_0\gg1$, this applies if $\gamma_e$ for the emitting electron is relatively low (see the second condition of the applicability of the LCFA). 
Let us compare the time scales of emission and electron acceleration for the parameters of interest here. 
During acceleration by the field, the electron gains energy $\sim mc^2$ at a time scale $1/\omega \atot\lesssim t_{\mathrm{rel}}\ll 1/\omega$. The characteristic time, at which it emits a photon, can be estimated from the LCFA rate: $\tem\sim \WCS^{-1}(\gamma_e(\tem),\chi_e(\tem))$ [see also Eq.~\eqref{tem_gem_cem} and Ref.\cite{mercuri2025growth}]. Assuming that the emission takes place at $\chi_e\lesssim 1$ (at $\chi_e\gg1$ the LCFA is typically valid), the low-$\chi_e$ asymptotic of Eq.~\eqref{eq:rate_CS_asympt} gives $\tem\sim 1/\sqrt{\alpha \atot}\omega$. Hence, $\tem/t_{\mathrm{rel}}\sim \sqrt{\atot/\alpha}\gg1$ for $\atot\sim 10^3$. Therefore, the electron gains high $\gamma_e$ during the phase of semi-classic motion before the instance of emission, which, in turn, can be safely treated within the LCFA. We note that the robustness of the LCFA should be verified with a different approach if $\omega\tem\gtrsim 1$. However, this case does not fall within the scope of the current work.
}

\CORR{
It should also be noted that the LCFA is known to fail in the infrared part of the radiation spectrum.\cite{DiPiazza2018_LCFA,di2019improved,ilderton2019extended} However, soft photons do not contribute to the formation of cascades, as such photons do not create pairs and do not cause significant recoil. Hence, the region of the parameter space where LCFA fails is irrelevant for modelling avalanche-type cascades. 
}

\CORR{
Finally, when higher-order processes in a cascade (such as tridents) are factorized, the spin states of intermediate particles should be treated as well. The common approach is taking spin-averaged probability rates for the first-order processes [given in Eqs.~\eqref{eq:rate_CS} and \eqref{eq:rate_BW}]. A more rigorous approach suggests using spin-resolved rates.\cite{king2013photon, seipt2021polarized} It is possible to include such rates explicitly in the kinetic description\cite{seipt2023kinetic} and simulation codes.\cite{zhao2023cascade,qian2023parametric} The main effect is the cascade growth rate dampening. However, simulations show that for background fields of optical frequency, calculations with spin-averaged rates provide a reliable result for the avalanche growth rate (we also assume that seed electron states are averaged over their initial states).\cite{king2013photon,seipt2021polarized} In our simulations with SMILEI,\cite{derouillat2018smilei} we rely on spin-averaged rates for the quantum processes. Still, it can be interesting to include spin dynamics in future works to identify its effect on final particle distributions in the parameter range of future experiments. 
}

\CORR{
A more rigorous derivation of all the listed approximations to modelling avalanches within the kinetic approach would be undoubtedly of great interest and should be done in future, for example, with the techniques developed in Ref.\cite{fauth2021collisional}
}

\section{\CORR{Dependence of the final particle spectra on the angular spread}}
\label{app:theta}
\begin{figure}
	\centering
	\includegraphics[width=\linewidth]{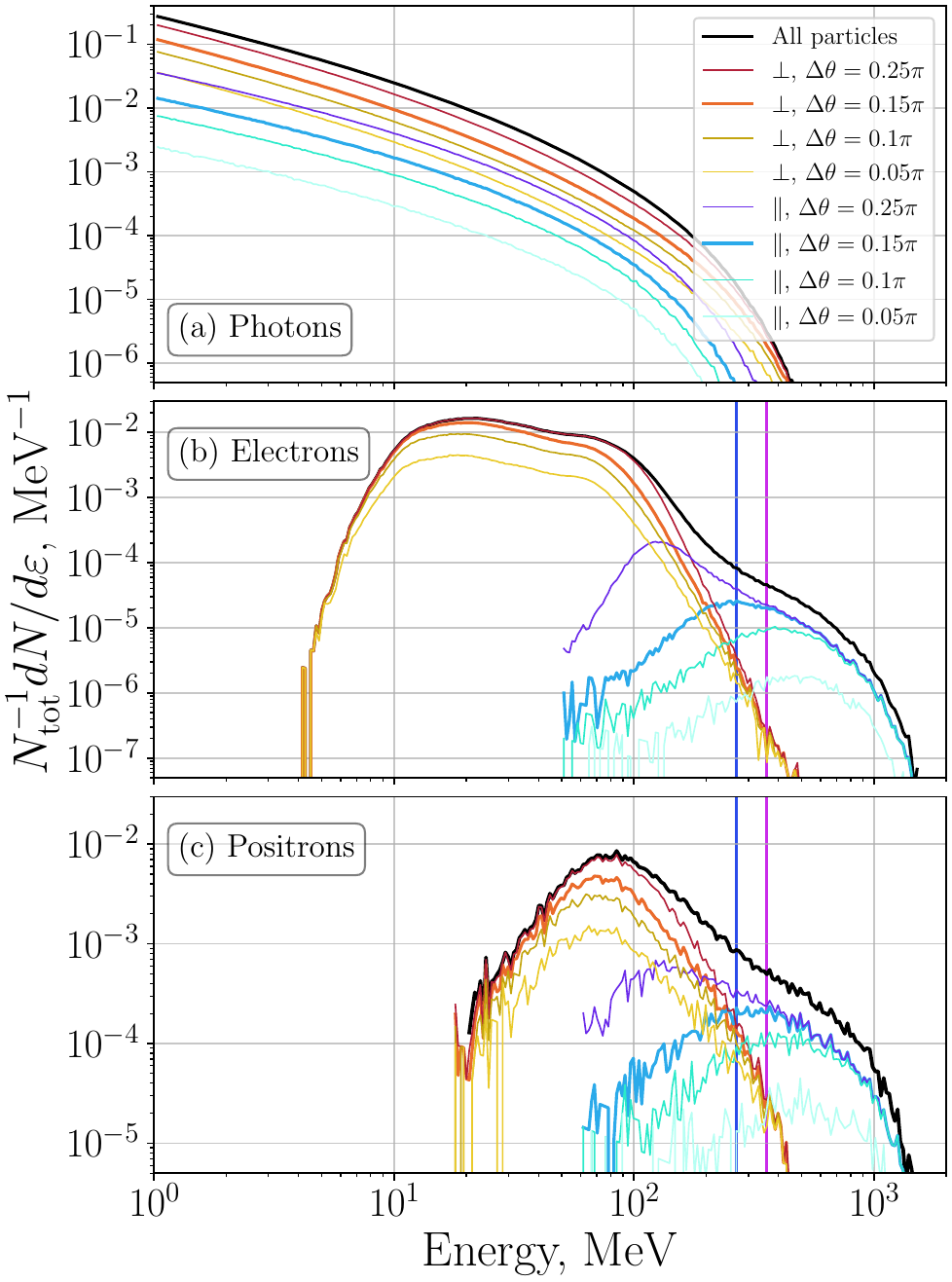}
	\caption{\CORR{The final particle spectra. The black, orange, and darker blue curves, as well as the vertical lines, are taken from Fig.~\ref{fig:spectra}. The spectra of particles propagating in the transverse direction $\pi/2-\Delta\theta<\theta<\pi/2+\Delta\theta$ (denoted ``$\perp$'' in the legend) are shown in warm colors for different values of the opening angle $\Delta\theta$. Similarly, curves in cold colors correspond to particles propagating longitudinally $0<\theta<\Delta\theta$ (denoted ``$\parallel$'' in the legend).}}
	\label{fig:spectra_delta_theta}
\end{figure}

\CORR{In Section \ref{sec:iiie}, we introduced the angular spread $\Delta\theta$ to separate the dynamics of particles propagating in the transverse direction $\pi/2-\Delta\theta<\theta<\pi/2+\Delta\theta$ and in the longitudinal direction $0<\theta<\Delta\theta$. As we discussed, particles propagating transversally are associated with the avalanche-type cascade, and one can study, for example, their spectra selectively, as shown in Fig.~\ref{fig:spectra}. In Section \ref{sec:iiie} and later in Fig.~\ref{fig:energy}, we set $\Delta\theta=0.15\pi$, as this value is convenient for illustration. For example, this can be seen from the green curve in Fig.~\ref{fig:spectra_angular}e showing the distribution of positrons in $\theta$. We associate the pronounced hump centered at $\theta=\pi/2$ to particles propagating directly from the interaction point, where the cascade is developing. The smaller peaks near $\theta\approx0$ and $\theta\approx\pi$ correspond to particles accelerated by each of the laser pulses, as they propagate after the impact. We find the spread $\Delta\theta=0.15\pi$ representative to discriminate between the longitudinal and transverse motion.}

\CORR{Here, we show the dependence of the final particle spectra on $\Delta\theta$, which is presented in Fig.~\ref{fig:spectra_delta_theta}. As one can see, the overall shape of the spectra is stable at varied $\Delta\theta$. In particular, the position of the high-energy cut-off for electrons and positrons propagating transversely matches for different $\Delta\theta$. At the largest spread, $\Delta\theta=0.25\pi$, the longitudinal distribution (the purple line) becomes noticeably wider in the lower energy. This motivates considering narrower angular spreads for the extraction of cleaner signals. Overall, our discussion and conclusions presented above are not sensitive to variation of $\Delta\theta$ within reasonable limits (namely, not too small so that there is enough statistics for a measurement, and not too large, so that the transverse and longitudinal motion are separated). }

\bibliography{lit}

\end{document}